\documentclass[10pt,prx,twocolumn,notitlepage,showpacs,superscriptaddress]{revtex4-1}
\usepackage{amsthm,amsmath,amsfonts,amssymb,verbatim, color}
\usepackage{graphicx}
\usepackage{bm}
\usepackage[colorlinks=true,citecolor=blue,linkcolor=blue,urlcolor=blue]{hyperref}
\usepackage{graphicx}
\usepackage{amsthm}
\usepackage{bbold}
\usepackage{wasysym}
\usepackage{enumerate}
\usepackage{color}
\usepackage{bbm}
\usepackage{xfrac}
\usepackage[dvips]{epsfig}
\usepackage[normalem]{ulem}

\newcommand{\ClebschG}[6]{\left[{#1\atop#2}{#3\atop#4}{#5\atop#6}\right]}

\newcommand{\ket}[1]{| #1 \rangle}
\newcommand{\SixJun}[6]{\left\{{#1\atop#2}{#3\atop#4}{#5\atop#6}\right\}}

\begin{document}
\title{Bulk Entanglement Spectrum in Gapped Spin Ladders}
\author{Raul A. Santos}
\affiliation{Department of Condensed Matter Physics, Weizmann Institute of Science, Rehovot, Israel; Department of Physics, 
Bar-Ilan University, Ramat Gan, 52900, Israel}
\author{Chao-Ming Jian}
\affiliation{Department of Physics, Stanford University, Stanford, California 94305, USA.}
\author{Rex Lundgren}
\affiliation{Department of Physics, The University of Texas at Austin, Austin, TX 78712, USA}

\begin{abstract}
We study the bulk entanglement of a series of gapped ground states of spin ladders,
representative of the Haldane phase. These ground states of spin $S/2$ ladders generalize the valence bond solid ground state. 
In the case of spin 1/2 ladders, we study a generalization of the Affleck-Kennedy-Lieb-Tasaki and Nersesyan-Tsvelik states and 
fully characterize the bulk entanglement Hamiltonian. In the case of general spin $S$ we argue that in the Haldane phase the bulk 
entanglement spectrum of a half integer ladder is either gapless or possess a degenerate ground state. For ladders with integer 
valued spin particles, the generic bulk entanglement spectrum should have an entanglement gap. Finally, we give an example of a 
series of trivial states of higher spin $S>1$ in which the bulk entanglement Hamiltonian is critical, signaling that the relation 
between topological states and a critical bulk entanglement Hamiltonian is not unique to topological systems.
\end{abstract}

%\pacs{71.10.Jm,71.10.Kt,75.50.Mm}

\maketitle

%%%%%%%%%%%%%%%%%%%%
\section{Introduction}
%%%%%%%%%%%%%%%%%%%%
Quantum entanglement has become a novel tool in the study of condensed matter systems due to its ability to reveal 
information about topological phases of matter, where there is no obvious order parameter \cite{kitaev2006,levin2006,jiang2012}. 
Entanglement between two subsystems, $A$ and $B$, can be characterized by the eigenvalues of the reduced density matrix of region $A$ (or equivalently $B$)
which is given by $\rho_A=\mathrm{tr}_B \rho$, where $\rho$ is the density matrix of the system in the ground state, i.e. the 
projector on the ground state. The eigenvalues of $\rho_A$ are called the entanglement spectrum (ES) \cite{PhysRevLett.101.010504}
and have been used to characterize topological order \cite{fidkowski2010,Chandran2011,qi2012}.

A recent trend in the study of the entanglement spectrum is to partition the system in ways other than the standard real-space 
bipartition. Examples include momentum-space partitions \cite{PhysRevLett.105.116805,PhysRevLett.113.256404,PhysRevLett.110.046806,2014arXiv1412.8612L,1742-5468-2014-7-P07022}, 
a partition between spin and orbital degrees of freedom \cite{Lundgren}, and an extensive real-space partition which is 
referred to as the bulk entanglement spectrum (BES) \cite{PhysRevLett.113.106801}. It has been 
argued that the BES of a topological state encodes information about the transition between
the topological and trivial phase within that system \cite{PhysRevLett.113.106801}. This connection has since been explored in 
quantum Hall states \cite{zhu2014,fukui2014,lu2015}, holographic and topological superconductors \cite{Belin2015,borchmann2014}, 
the metal-insulator transition \cite{Vijay2015} and symmetry protected topological states \cite{PhysRevB.90.085137,PhysRevB.90.075151,Santos2015,lu2015b}.

In this work, we investigate the BES of spin ladders of arbitrary spin. The states that we discuss
are generalizations of the valence bond solid (VBS) ground state. In one dimension, the VBS state is the exact
ground state of the Affleck-Kennedy-Lieb-Tasaki (AKLT) spin chain model and its higher spin generalizations \cite{Xu2008}. These states 
realize a symmetry protected topological phase when the spin of the particles, $S$, in the chain is an odd integer \cite{PhysRevB.85.075125}.
For $S$ even, the system is topologically trivial. The $S=1$ AKLT spin chain does not have a long range order but instead
posses nonlocal ordering characterized by a string order parameter \cite{Nijs1986}, which is representative of the Haldane phase 
\cite{PhysRevLett.50.1153,Kennedy1992}. Recently, it has been argued that the BES in a topologically non trivial state is critical, 
meaning that the low lying entanglement spectrum levels are either degenerate or gapless (in the thermodynamical limit) 
\cite{PhysRevLett.113.106801}. This has been shown explicitly for the AKLT ground state and some generalizations 
\cite{PhysRevLett.113.106801,PhysRevB.90.085137,PhysRevB.90.075151,Santos2015}. In these works, it has been shown that the BES 
of the $S=1$ VBS ground state is described by a conformal theory of central charge $c=1$.

We discuss the connection between topologically trivial/nontrivial states and the critical/non-critical
BES in spin ladders. As pointed out in Refs.~\cite{Dagotto1992} and~\cite{gogolin2004bosonization}, the Haldane phase can be	
realized by coupling two identical spin $1/2$ chains. It is then natural to ask whether the BES of the Haldane phase realized in 
spin ladders is still critical. We find that for spin-ladder ground states representative of the Haldane phase, the BES is indeed 
gapless. More specifically, we analyze the most general spin 1/2 ladder ground state defined by a matrix product state (MPS) of 
bond dimension two, invariant under time reversal symmetry. This state contains the AKLT groundstate as a limit. We carry out a 
complete analysis of this generalized state, computing the two point correlation functions along the legs and the rungs. We 
show that degeneracies in the transfer matrix make the BES gapless, and compute the bulk entanglement Hamiltonian (BEH)
for three inequivalent extensive partitions. In most cases, the BEH for all these partitions corresponds to an XYZ effective spin 1/2 model, 
where the coupling constants depend on the specific partition. For all partitions considered, the BEH becomes critical whenever 
the transfer matrix has degenerate eigenvalues.
 
We also consider a spin $S/2$ $SU(2)$ symmetric ladder, which is introduced by means of its MPS representation and is adiabatically 
connected to the VBS ground state. We show that it is possible to extract the BEH (which is an operator acting on virtual spins 
of size $S/2$) perturbatively from the MPS representation. Tracing every other rung of the ladder, we find that the BEH corresponds
to a ferromagnetic effective Hamiltonian of a spin $S/2$ chain. Tracing every two other rungs, the BEH corresponds to a
antiferromagnetic Hamiltonian. Thus, by the Haldane conjecture \cite{PhysRevLett.50.1153}, we argue that generically the BEH in 
these systems is gapped for $S$ even and gapless for $S$ odd. A special deformation is then introduced that shows the 
non-universality of the bulk entanglement Hamiltonian \cite{Chandran2014}. This deformation links the spin $S$ VBS ground state 
with an $SU(N)$ symmetric state (with $N=S+1$) in the same phase, keeping finite the correlation length along the deformation. 
It allows us to obtain exactly the BES by mapping the R\'{e}nyi entropy to the partition function of a two dimensional Potts 
model at criticality, where the critical model represents a first order phase transition. We argue that this is a 
signal of a dimerization transition in the physical ground state.

Our paper is organized as follows. In Sec.~\ref{sec:spin-half}, the BES of spin 1/2 ladders is calculated. In 
Sec.~\ref{sec:spinSladder}, we discuss spin $S/2$ ladders and their bulk entanglement properties. In Sec.~\ref{sec:NON_UNI}, 
we investigate the non-universal properties of the BES. Finally, in Sec.~\ref{sec:CON}, we present our conclusions. Some 
technical information and details are presented in the Appendix.

%%%%%%%%%%%%%%%%%%%%
\section{$S=1/2$ Spin Ladders}\label{sec:spin-half}
%%%%%%%%%%%%%%%%%%%%
In this section, we consider the BES of gapped spin-half ladders. The ground state wave function of such states can be described 
in terms of MPS \cite{Perez-Garcia2007}.
We focus on MPS (of bond dimension two) that generalize the AKLT \cite{PhysRevLett.59.799,ALKT_LONG} ground state. More 
specifically, we consider the 
following $SO(3)$ symmetric ground state MPS wave function of the ladder of the form
\begin{equation}\label{NT_MPS}
|\psi_0\rangle=\mathrm{tr}(g_1(u)g_2(\tilde{u})\cdots g_{2N-1}(\tilde{u})g_{2N}(u)),
\end{equation}
where
\begin{equation}
g_{i}(u)=u\mathbb{I}_{2\times2}|s\rangle+\sqrt{2}\sigma^{+}|+\rangle+\sigma^z|0\rangle-\sqrt{2}\sigma^{-}|-\rangle,
\end{equation}
where $\mathbb{I}_{2\times2}$ is the $2\times2$ identity matrix. The Pauli matrices are given by
\begin{equation}\label{pauli_m}
\sigma^+ =
 \begin{pmatrix}
  0 & 1 \\
  0 & 0 
 \end{pmatrix},~
\sigma^- =
 \begin{pmatrix}
  0 & 0 \\
  1 & 0 
 \end{pmatrix},~
\sigma^z =
 \begin{pmatrix}
  1 &  0 \\
  0 & -1
 \end{pmatrix}.
\end{equation}
Here $|s\rangle$ labels the singlet state composed of two spin 1/2 sitting in a rung and $|+\rangle,|0\rangle,|-\rangle$ label 
$m=1,0,-1$ in the triplet state of one rung. 

\begin{figure}[tb]
	\centering
		\includegraphics[width=0.85\linewidth]{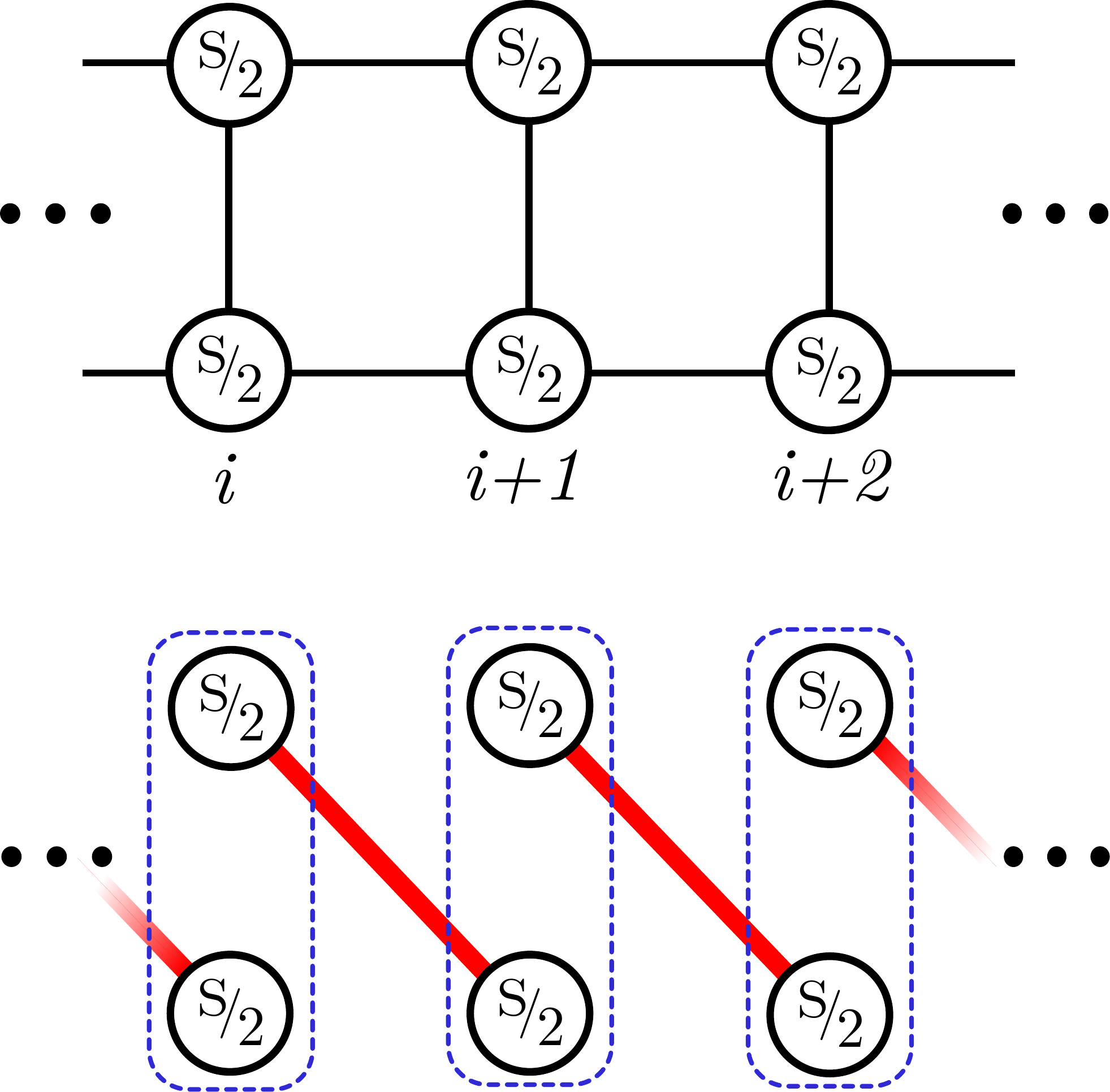}
	\caption{(Color online) Top: A ladder formed by spins $S/2$. In our notation, on each site, $i$, there are two
	$S/2$ particles corresponding to the top and bottom particle. Bottom: The AKLT ground state is obtained
	by projecting the states in each site to the symmetric subspace (of total spin $S$) and creating singlets
	between top-bottom pairs at different sites. Symmetrization at each site is denoted by blue dashed boxes. 
	Singlets correspond to red lines connecting pairs of particles.
	The NT phase is obtained without symmetrizing the rungs and it consists of dimer coverings of the ladder. 
	In this work, we have chosen one of the two possible dimer coverings.}\label{fig:Spin_S_ladder}
\end{figure}

This wave function can be thought as a generalization of the AKLT spin-1 chain where now each site hosts two real spin 1/2 particles and we allow for the existence of singlet states on each site (See Fig. \ref{fig:Spin_S_ladder} with $S=1/2$). The VBS wave function corresponds to $u=\tilde{u}=0$ \cite{PhysRevLett.59.799,ALKT_LONG}. At this point the MPS agrees with the AKLT ground state which belongs to the Haldane phase, with a non-zero string order parameter and fractionalized spin-half edge states \cite{Kennedy1992}.  For $u=-\tilde{u}=1$, the wave function (\ref{NT_MPS}) represents a spin-liquid phase \cite{PhysRevLett.80.2709}, which was originally proposed by Nersesyan and Tsvelik \cite{PhysRevLett.78.3939}. We refer to this phase as the NT phase. Both the AKLT and NT ground states are depicted in the bottom of Fig.~\ref{fig:Spin_S_ladder}.

The Haldane and NT phases are thermodynamically indistinguishable from each other, however their correlation functions 
differ drastically \cite{PhysRevLett.78.3939}. In the Haldane phase the spinon excitations of the legs of the ladders confine and form triplet (magnon) 
and singlet excitations, with a triplet and singlet mass gap, $m_t$ and $m_s$ respectively. There is a coherent peak in the 
dynamical spin susceptibility, $\chi''(\omega,q)$, at $q=\pi$ and $\omega=m_t$ due to the triplet excitations. In contrast, the 
NT phase has no coherent excitations \cite{PhysRevLett.78.3939}. The entanglement properties of these two phases also 
have similarities and differences \cite{Lundgren}. For a bipartite partition, there is a protected two-fold 
degeneracy in the entanglement spectrum for both phases. For a partition between the legs of the ladder, the entanglement spectrum 
of the Haldane phase is described by a conformal field theory (CFT) with a 
central charge $c=1$, while the entanglement spectrum for the NT phase consists only of two entanglement 
levels (when one takes a certain linear combination of the two possible dimer coverings). Combining the results of both of these partitions allows one to tell the difference between these two phases.

We now generalize Eq.~(\ref{NT_MPS}) further. This will allow us to characterize the different phases of the entanglement Hamiltonian as a function of the MPS wave function.
It is defined by
\begin{equation}\label{Ground_state_ladder}
 |G\rangle=\mathrm{tr}(M_1M_2\cdots M_{N-1}M_{N}),
\end{equation}
with $M_j=\sum_{J=0}^1\sum_{m=-J}^J M(J,m)|J,m\rangle_j.$
In this section, we concentrate on the set of $2\times 2$ matrices $M(J,m)$, which are given by 
\begin{eqnarray}\label{MPS_spin_half_sol}
 M(0,0) &=&\frac{\alpha+\beta}{\sqrt{2}}\mathbb{I},\quad M(1,0) =\frac{\alpha-\beta}{\sqrt{2}}\sigma_z,\\\nonumber
 M(1,1) &=& -(\delta\sigma^++\gamma\sigma^-),
\quad
M(1,-1)=\gamma\sigma^++\delta\sigma^-,
\end{eqnarray}
where $\alpha,\beta,\gamma,\delta$ are real numbers and $|J,m\rangle_j$ is the state on the rung, composed of two spin 1/2 particles.
Eq. (\ref{MPS_spin_half_sol}) is the simplest nontrivial set of $2\times 2$ matrices leading to a time reversal invariant ground state $|G\rangle$.
Eq. (\ref{Ground_state_ladder}) reproduces Eq. (\ref{NT_MPS}) for $(\alpha,\beta,\gamma,\delta)=(u+1,u-1,0,2)$
or $\gamma\leftrightarrow\delta$.

It will prove useful to express Eq. (\ref{Ground_state_ladder}) as function of the spin 1/2 particles explicitly.
We label the rungs of the ladder by $i\in\{-\infty,\infty\}$ and the legs by $1$ (upper leg) and
$2$ (lower leg). The MPS is explicitly
\begin{equation}\label{MPS_spin_half}
 M_j=\sum_{a,b,b'=-}^{+} M^{[a,b]}\epsilon_{b,b'}|a_j,b'_{j}\rangle,
\end{equation}
where $|a_j,b_j\rangle$ is a compact notation for $|a\rangle_{1,j}\otimes|b\rangle_{2,j}$.
$|\pm\rangle_{\eta,j}$ is the eigenstate of $s^z_j$ of a spin 1/2 particle located on the $i$th rung and on the $\eta$th leg 
with eigenvalue $\pm$ (see Fig. \ref{fig:Spin_S_ladder}). The matrix $\epsilon$ is an antisymmetric matrix such that $\epsilon_{+-}=-\epsilon_{-+}=1$. 
The matrices $M^{[a,b]}$ are in turn

\begin{eqnarray}\label{MPS_sim}
M^{[++]} &=&\begin{pmatrix}
  \alpha & 0 \\
  0 & \beta 
 \end{pmatrix},\,\,
M^{[--]} =
 \begin{pmatrix}
  \beta & 0 \\
  0 & \alpha 
 \end{pmatrix},\\\nonumber
M^{[-+]} &=&
 \begin{pmatrix}
  0 &  \gamma \\
  \delta & 0
 \end{pmatrix},\,\,
M^{[+-]}=
  \begin{pmatrix}
  0 &  \delta \\
  \gamma & 0
 \end{pmatrix}.
\end{eqnarray}
This generalized MPS is the groundstate of a local parent Hamiltonian \cite{Perez-Garcia2007,nakahara2012frontiers}.

%%%%%%%%%%%%%%%%%%%%
\subsection{On-site symmetries of MPS}
The symmetries of the MPS dictate the symmetries of the ground state. These symmetries
can be used to classify the symmetry protected topological phases in one dimension \cite{Schuch2011}.
We now discuss the symmetries present in Eq. (\ref{Ground_state_ladder}) which include time reversal, 
$\mathbb{Z}_2\times \mathbb{Z}_2$, and rotation symmetry.

%%%%%%%%%%%%%%%%%%%%
\subsubsection{Time Reversal Symmetry}
%%%%%%%%%%%%%%%%%%%%

Under time reversal $\mathcal{T}$, 
$M_j$ changes as $\mathcal{T}M_j=\sum_{J,m} M(J,-m)^*(-1)^{J+m}|J,m\rangle$. If the ground state is invariant under time 
reversal, the matrices $M(J,m)$ must transform as
\begin{equation}\label{TR_cond}
 M(J,-m)^*(-1)^{J+m}=U^{-1}M(J,m)U,
\end{equation}
with $M(J,m)^*$ the complex conjugate of $ M(J,m)$. Here $U$ is a $2\times 2$ matrix realizing a projective representation 
of $\mathcal{T}$. The projective representation acts on the matrix indices of $M(J,m)$ and is independent of $J$ and $m$. It is easy to see that for the matrices $M(J,m)$ defined by Eq. (\ref{MPS_spin_half_sol}), the matrix $U$ that realizes the projective representation of time reversal symmetry is simply $U=i(\sigma^--\sigma^+)=\sigma_y$. The simplest nontrivial solution of Eq. (\ref{TR_cond}) is given by Eq. (\ref{MPS_spin_half_sol}).
%%%%%%%%%%%%%%%%%%%%
\subsubsection{$\mathbb{Z}_2\times \mathbb{Z}_2$ and rotation symmetry}
%%%%%%%%%%%%%%%%%%%%

The $\mathbb{Z}_2\times \mathbb{Z}_2$ symmetry that we consider is generated by the action of any pair of 
the operators $\hat{O}_{x,y,z}=e^{i\pi \hat{S}_{x,y,z}}$, with $\hat{S}_a$ the total spin operator in the rung. Eq. (\ref{MPS_spin_half_sol}) is symmetric under the action of all $O_{x,y,z}$ operators, with $\hat{O}_aM_j=\sigma_aM_j\sigma_a$,
where $\sigma_a$ $(a=x,y,z)$ are Pauli matrices. Eq. (\ref{MPS_spin_half_sol}) also has rotation symmetry around the $z$ axis for vanishing $\delta$ or $\gamma$. Full rotational symmetry is achieved for $\delta$ (or $\gamma)=0$ and $\alpha-\beta=\gamma$ (or $\delta)$.
%%%%%%%%%%%%%%%%%%%%
\subsection{Ground State correlation functions}
\begin{figure}[h!]
	\centering
		\includegraphics[width=0.9\linewidth]{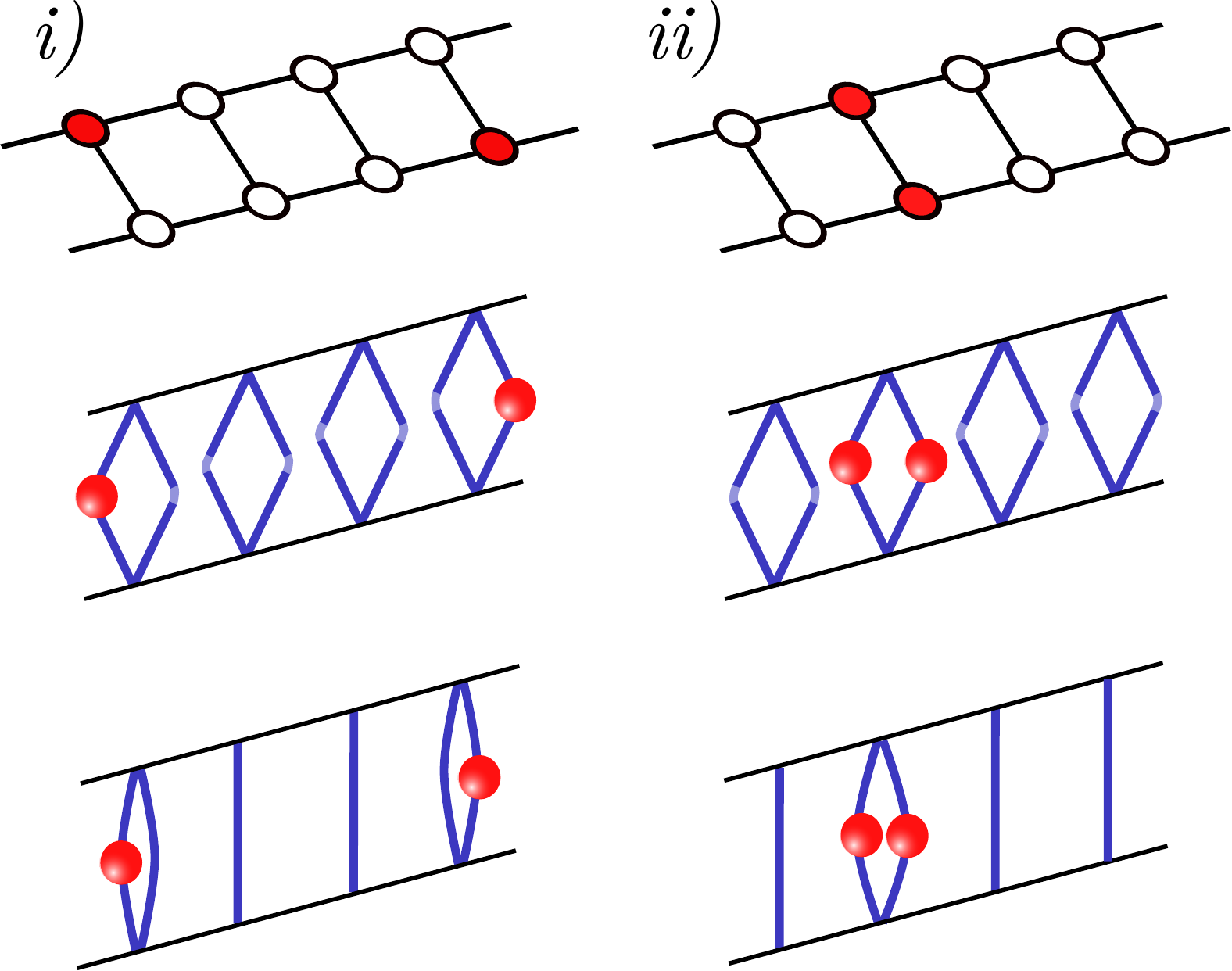}
	\caption{(Color online) i) Correlation function of operators in different rungs. In terms of the one dimensional transfer matrix,
	the correlation function is obtained by acting with the operators (red spheres) in the spin states and then 
	contracting the physical states (blue lines) ii) Correlation of operators in the same rung.
		}\label{fig:Corr}
\end{figure}

%%%%%%%%%%%%%%%%%%%%
The MPS representation of the ground state wave function in gapped one dimensional systems makes the computation of correlation functions straightforward 
(see Fig. (\ref{fig:Corr})) and makes use of the one dimensional transfer matrix 
$(\tilde{T}_{r})_{ab}^{cd}=\sum_{j,m}M(j,m)_{ab}\otimes M^*(j,m)^{cd}$
\begin{equation}\label{TM_rung}
\tilde{T}_{r}= 
\begin{pmatrix}
  \alpha^2+\beta^2 & 0 & 0 & \delta^2+\gamma^2\\
  0 & 2\alpha\beta & 2\gamma\delta & 0 \\
 0  & 2\gamma\delta & 2\alpha\beta  & 0 \\
 \delta^2+\gamma^2  & 0 & 0  & \alpha^2+\beta^2
 \end{pmatrix}.
\end{equation}
We concentrate on two types of spin-spin correlations, the correlations of operators acting in the same rung, and 
the correlation of operators acting on different rungs. Note that the last case includes the correlation between
different sites. As we will see next, the correlation lengths for the different correlators are connected with
the eigenvalues of the transfer density matrix

%%%%%%%%%%%%%%%%%%%%
\subsubsection{Correlations within a rung}
%%%%%%%%%%%%%%%%%%%%
The correlation function of two operators, $\mathcal{A}$,$\mathcal{B}$, acting on the $i$th rung is 
\begin{equation}\label{DCF_rungs}
 \langle \mathcal{A}_i\mathcal{B}_{i}\rangle=\frac{{\rm tr}\langle 0|T(\mathcal{A}_i,\mathcal{B}_{i})|0\rangle}{\alpha^2+\beta^2+\gamma^2+\delta^2}.
\end{equation}
where we have assumed that the ladder is infinite. The vector $|0\rangle$ corresponds to the eigenvector of Eq. (\ref{TM_rung}) with the largest eigenvalue. The matrix $T(\mathcal{A}_i,\mathcal{B}_{i})$ is given by

\begin{equation}
 T(\mathcal{A}_i,\mathcal{B}_i)=\sum_{\substack{j_i,m_i\\j'_i,m'_i}}\langle j_i,m_i|\mathcal{A}_i\mathcal{B}_i|j'_i,m'_i\rangle M(j_i,m_i)\otimes M^*(j'_i,m'_i).
\end{equation}

We denote the spin operators by $\hat{s}_{\eta,i}^a$.
Two spins on the same rung have the following correlation functions,
\begin{equation}\label{CF_rung}
 \langle s_{1,i}^as_{2,i}^b\rangle=\delta^{ab}\frac{r_a}{4(\alpha^2+\beta^2+\gamma^2+\delta^2)}.
\end{equation}
Here, $r_x=-2(\alpha\beta+\gamma\delta)$, $r_y=2(\gamma\delta-\alpha\beta)$
and $r_z=\gamma^2+\delta^2-\alpha^2-\beta^2$.
%%%%%%%%%%%%%%%%%%%%
\subsubsection{Correlations at different rungs}
%%%%%%%%%%%%%%%%%%%%
The correlation function of two operators $\mathcal{A}$ and $\mathcal{B}$ (assuming $L\neq 0$) in an infinite ladder is
\begin{equation}
 \langle \mathcal{A}_i\mathcal{B}_{i+L}\rangle=\frac{\langle 0|T(\mathcal{A}_i)\tilde{T_r}^{L-2}T(\mathcal{B}_{i+L})|0\rangle}{(\alpha^2+\beta^2+\gamma^2+\delta^2)^L}.
\end{equation}
$T(\mathcal{O}_i)$ becomes
\begin{equation}
 T(\mathcal{O}_i)=\sum_{\substack{j_i,m_i\\j'_i,m'_i}}M(j_i,m_i)\langle j_i,m_i|\mathcal{O}_i|j'_i,m'_i\rangle\otimes M^*(j'_i,m'_i).
\end{equation}

\noindent Using the Eq. (\ref{MPS_spin_half_sol}), we obtain the pair correlation function of spin operators,

\begin{equation}
 \langle \hat{s}_{\eta,i}^a\hat{s}_{\eta',i+L}^b\rangle=\delta^{ab}\bar{\phi}^a_\eta\phi^a_{\eta'}e^{-L/\xi_a}\chi_a^L,
\end{equation}

\noindent with $\phi^x,\phi^y$, and $\phi^z$ being

\begin{equation}
 \phi^x=\frac{1}{2(\alpha\beta+\gamma\delta)}\begin{pmatrix}
  \alpha\gamma+\beta\delta  \\
  -(\alpha\delta+\beta\gamma)
 \end{pmatrix},
\end{equation}

\begin{equation}
 \phi^y=\frac{1}{2(\alpha\beta-\gamma\delta)}\begin{pmatrix}
  \alpha\gamma-\beta\delta  \\
  \alpha\delta-\beta\gamma 
 \end{pmatrix},
\end{equation}

\noindent and 

\begin{equation}
 \phi^z=\frac{1}{2(\alpha^2+\beta^2-\gamma^2-\delta^2)}\begin{pmatrix}
  \alpha^2-\beta^2+\gamma^2-\delta^2  \\
  \beta^2-\alpha^2+\gamma^2-\delta^2
 \end{pmatrix},
\end{equation}

\noindent while $\bar{\phi}^a=-\tau\phi^a$, where $\tau=\begin{pmatrix} 0&1\\1&0\end{pmatrix}$. The correlation lengths
$\xi_a$, $(a=x,y,z)$ are explicitly

\begin{equation}\label{Corr_lenght_x}
\frac{1}{\xi_x}=\ln\left(\frac{2|\alpha\beta+\gamma\delta|}{\alpha^2+\beta^2+\gamma^2+\delta^2}\right),
\end{equation}

\begin{equation}\label{Corr_lenght_y}
\frac{1}{\xi_y}=\ln\left(\frac{2|\alpha\beta-\gamma\delta|}{\alpha^2+\beta^2+\gamma^2+\delta^2}\right),
\end{equation}

\begin{equation}\label{Corr_lenght_z}
\frac{1}{\xi_z}=\ln\left(\frac{|\alpha^2+\beta^2-\gamma^2-\delta^2|}{\alpha^2+\beta^2+\gamma^2+\delta^2}\right).
\end{equation}

\noindent Finally, ferromagnetic-antiferromagnetic order of the correlation functions is dictated by $\chi_a$, which 
corresponds respectively to $\chi_x={\rm sgn}(\alpha\beta+\gamma\delta)$, $\chi_y={\rm sgn}(\alpha\beta-\gamma\delta)$ and 
$\chi_z={\rm sgn}(\alpha^2+\beta^2-\gamma^2-\delta^2)$, with ${\rm sgn}(x)=x/|x|$ the sign function.

%%%%%%%%%%%%%%%%
\subsection{Bulk entanglement}
%%%%%%%%%%%%%%%%%%%%%

In this section, we analyze the bulk entanglement of the spin ladder wave function $|G\rangle$ defined in 
Eq. (\ref{Ground_state_ladder}). In general, to obtain the entanglement of a given state, we need to define two
complementary sets of sites which we denote by $A$ and $A^c$. Together, these two sets contain all sites in the lattice. The density matrix that describes the ground state is $\rho=|G\rangle\langle G|$ (for normalized $|G\rangle$). Tracing over one of the subsets, say $A^c$, we obtain the reduced density matrix, $\rho_A={\rm tr}_{A^c}\rho$. $\rho_A$ fully characterizes the entanglement between the regions $A$ and its complement $A^c$. If the boundary between $A$ and $A^c$ covers the whole lattice, we then have access to the entanglement in the bulk of the system \cite{PhysRevLett.113.106801}. There are of course an infinite number of ways of defining two complementary sets whose boundary traverses the whole lattice. We
investigate the simplest of them, invariant under lattice translations.
%%%%%%%%%%%%%%%%%%%%
\subsubsection{Tracing every other rung}\label{sec:bes}
\begin{figure}[h!]
	\centering
		\includegraphics[width=0.9\linewidth]{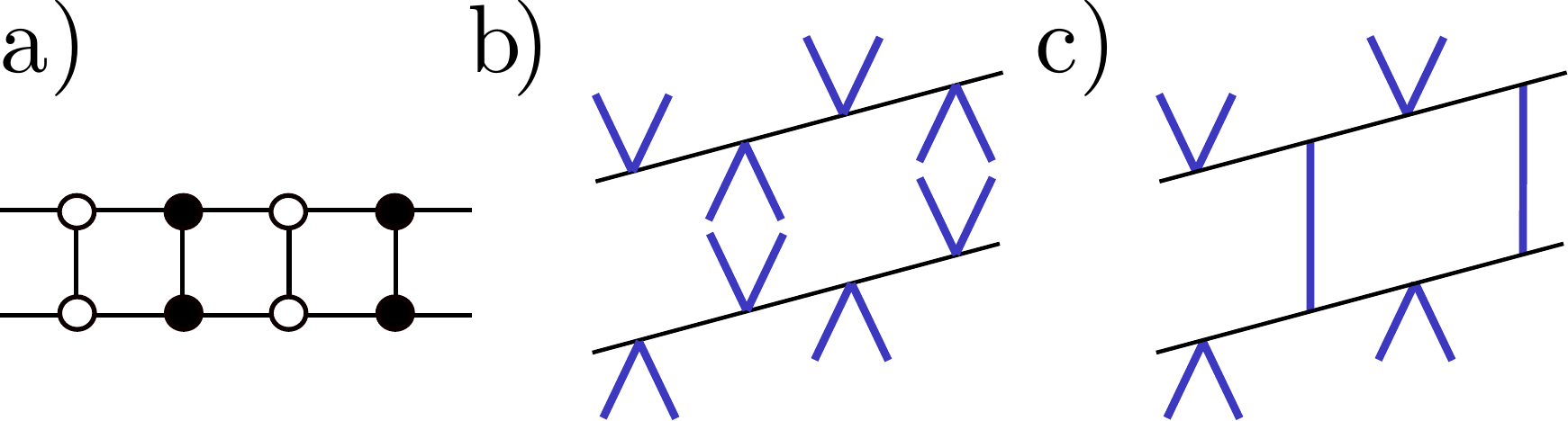}
	\caption{(Color online) a) Neighboring rungs belong to different subsets, represented  here by white 
	and black dots. b) Tracing the physical states of the sites in $A^c$. To obtain the reduced density matrix, we need to 
	trace the states in the sites in $A^c$, this means contracting the corresponding spin states in those sites. 
	c) Graphical description of $\rho_A$. After tracing the sites in $A^c$, the partial density matrix becomes an operator 
	the Hilbert space associated with the sites in $A$ }\label{fig:rung_partition}
\end{figure}

%%%%%%%%%%%%%%%%%%%%
We assign the sites in consecutive rungs to different sets, $A$ and $A^c$ (see Fig. (\ref{fig:rung_partition}a)). 
This partition allows us to investigate the entanglement between rungs in the ladder. The BES of this partition in the AKLT spin 1 point has been studied in several works \cite{PhysRevB.90.075151,PhysRevB.90.085137,Santos2015}. The main result of these works is that the bulk entanglement Hamiltonian is described by a $c=1$ CFT. 

\begin{figure}[h!]
	\centering
		\includegraphics[width=0.8\linewidth]{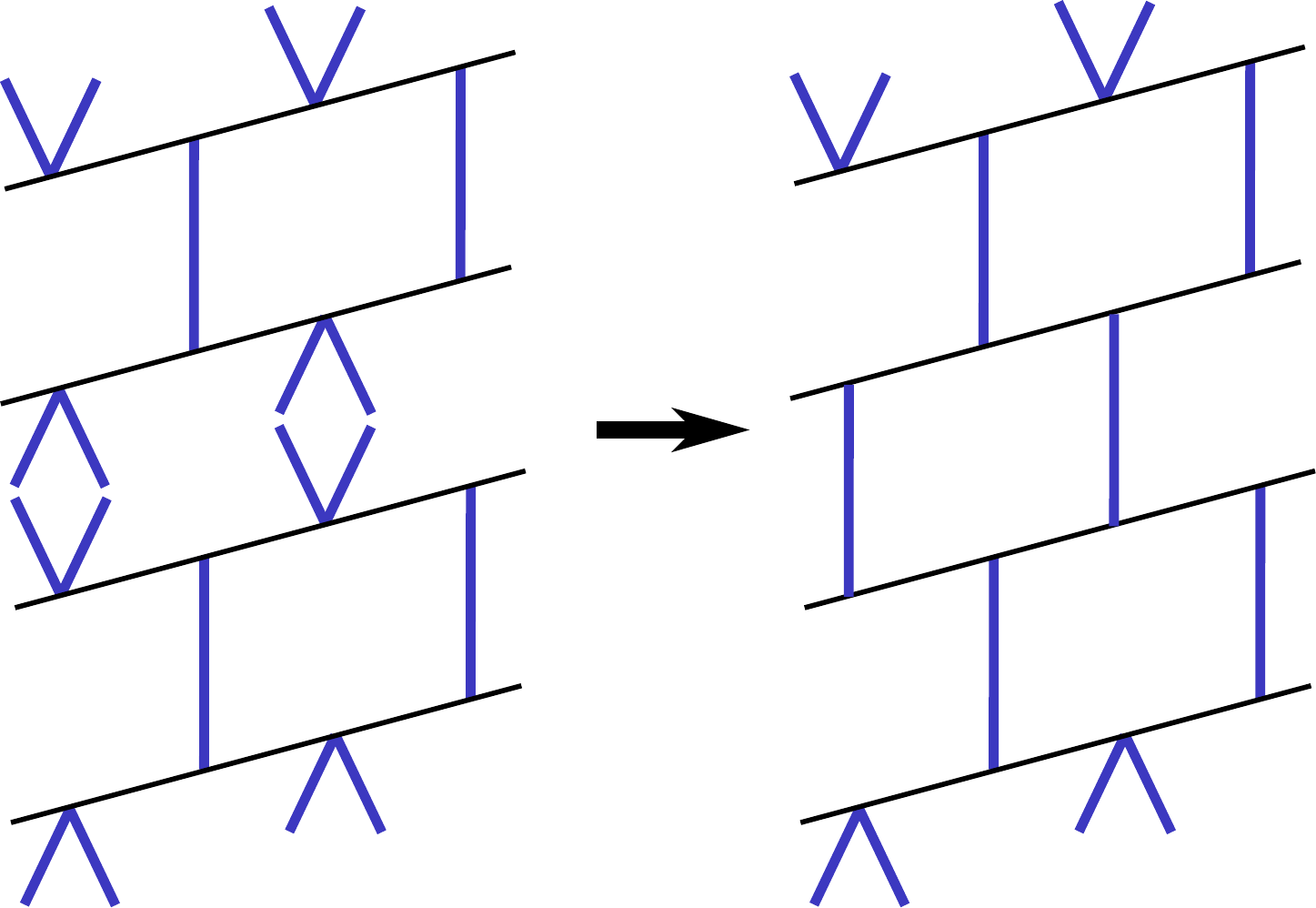}
	\caption{Staking $\rho_A$ vertically $n$-times an contracting the indices of the spin states, we obtain a classical partition function of width $N$ (the length of the ladder) and
	height $n$.}\label{fig:class_part_rung}
\end{figure}

In the following discussion, we will make use of the graphical representation of the MPS, already presented in the previous
section for the computation of correlation functions. For this partition, $\rho_A$ has a graphical form given by Fig. (\ref{fig:rung_partition}c), which is an operator that acts in the Hilbert subspace generated by the sites in $A$. The R\'{e}nyi entropy, $\mathcal{S}_n\sim {\rm tr}\rho_A^n$, can 
be obtained by stacking $n$ of these objects and contracting the corresponding spin indices, as seen in Fig. (\ref{fig:class_part_rung}). 
The result is that the R\'{e}nyi entropy is represented in terms of a classical partition function \cite{Santos2015}. For the spin ladder
ground state $|G\rangle$ (Eq. (\ref{Ground_state_ladder})), with periodic boundary conditions, the R\'{e}nyi entropy becomes the partition function of an eight vertex model on a torus \cite{PhysRevB.90.085137,Santos2015,baxter2013exactly}.

\begin{figure}[h!]
	\centering
		\includegraphics[width=0.9\linewidth]{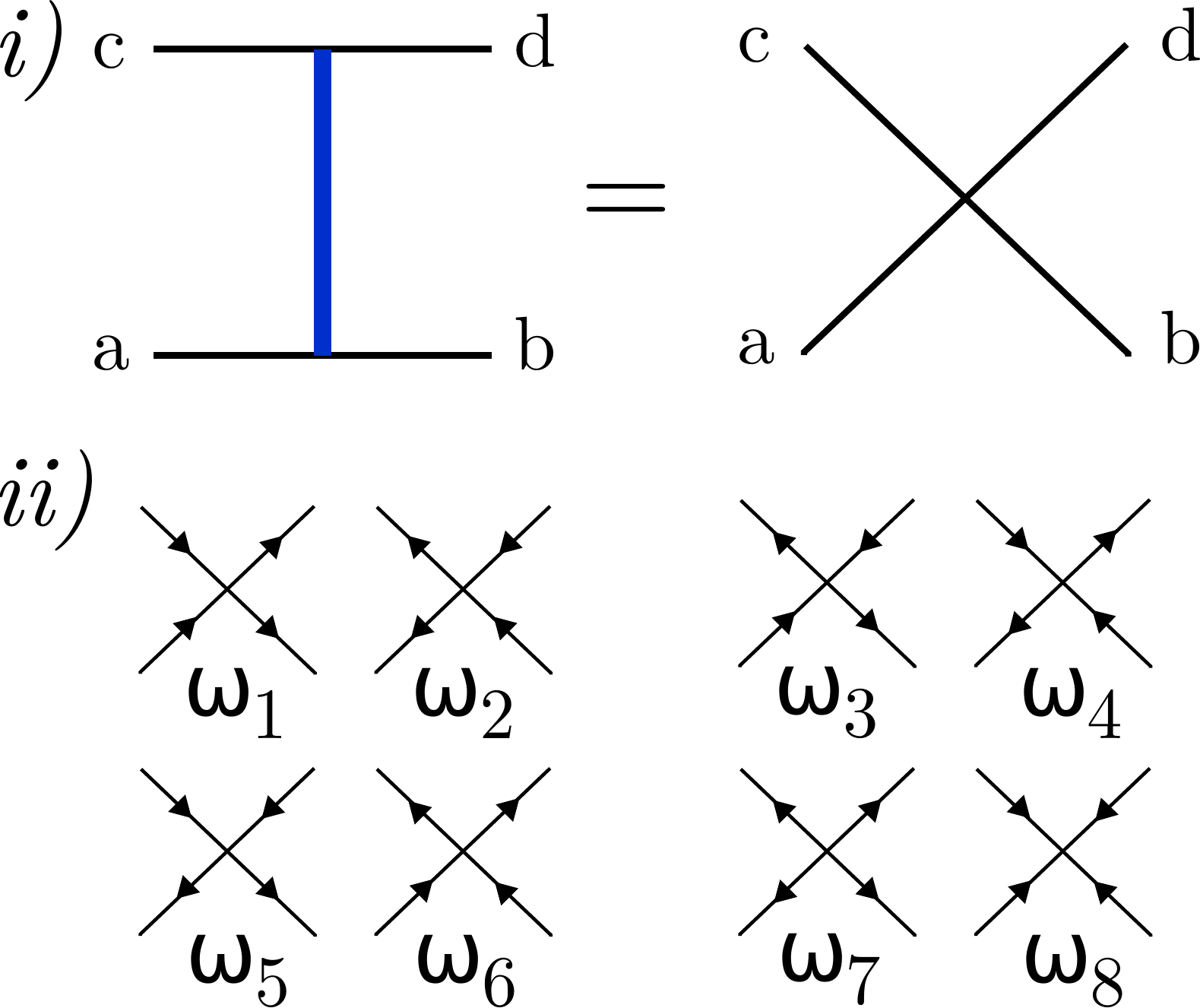}
	\caption{i) The one dimensional transfer matrix $\tilde{T}$ as an arrow configuration. $\tilde{T}$ generates the two
	dimensional partition function that determines the R\'{e}nyi entropy. The mapping to the eight vertex model is 
	follows from transforming the matrix $\tilde{T}$ into an arrow configuration. ii) Arrow configurations for
	the eight vertex model. In our convention, arrows pointing away from the crossing point are denoted by '$+$'.
	Arrows pointing to the crossing point correspond to '$-$'.
	The Boltzmann weights for these configurations are defined by the corresponding entry in the
	one dimensional transfer matrix $\tilde{T}$}\label{fig:TM_rungs}
\end{figure}

More concretely, the R\'{e}nyi entropy for this rung partition becomes 
\begin{equation}\label{Renyi_def}
 \mathcal{S}_n=\ln{\rm tr} \frac{\rho^n}{(1-n)}=\frac{\ln(Z_{rung}(\alpha,\beta,\gamma,\delta))-n\ln(Z_0)}{1-n},
\end{equation}
 with $Z_0=(\alpha^2+\beta^2+\gamma^2+\delta^2)$. Here, $Z_{rung}$ corresponds to the partition function of the eight vertex model
generated by the one dimensional transfer matrix (Eq. (\ref{TM_rung})) (see also Fig. (\ref{fig:TM_rungs}i)). The matrix, Eq. (\ref{TM_rung}),
defines the Boltzmann weights as
\begin{eqnarray}\nonumber
\omega_1=(\tilde{T_r})_{-+}^{-+}̣̣&=&\omega_2= (\tilde{T_r})_{+-}^{+-}̣̣ =\delta^2+\gamma^2,\\\nonumber
\omega_3=(\tilde{T_r})_{-+}^{+-}̣̣&=&\omega_4= (\tilde{T_r})_{+-}^{-+}̣̣ =2\delta\gamma,\\
\omega_5=(\tilde{T_r})_{++}^{--}̣̣&=&\omega_6= (\tilde{T_r})_{--}^{++}̣̣ =2\alpha\beta,\\\nonumber
\omega_7=(\tilde{T_r})_{++}^{++}̣̣&=&\omega_8= (\tilde{T_r})_{--}^{--}̣̣ =\alpha^2+\beta^2.
\end{eqnarray}

\noindent The relationship between the Boltzmann weights and the arrow configurations that define the eight vertex model are 
presented in Fig. (\ref{fig:TM_rungs}ii).

For a matrix of Boltzmann weights $\tilde{T}$ of the form
\begin{equation}
\tilde{T}= 
\begin{pmatrix}
  a & 0 & 0 & d\\
  0 & c & b & 0 \\
 0  & b & c  & 0 \\
 d & 0 & 0  & a
 \end{pmatrix},
\end{equation}
\noindent  we define two important quantities
\begin{equation}
\Delta=\frac{a^2+b^2-c^2-d^2}{2(ab+cd)},\quad \Gamma=\frac{ab-cd}{ab+cd}.
\end{equation}
The eight vertex model is critical for $\Delta=\pm1$ \cite{baxter2013exactly}. After tracing every other rung we have 
\begin{eqnarray}\label{Delta_rung}
 \Delta_{rung}&=&\frac{(\alpha^2-\beta^2)^2-(\gamma^2-\delta^2)^2}{4((\alpha+\beta)^2\gamma\delta+(\gamma-\delta)^2\alpha\beta)},\\
 \Gamma_{rung}&=&\frac{(\alpha^2+\beta^2)\gamma\delta-(\gamma^2+\delta^2)\alpha\beta}{(\alpha^2+\beta^2)\gamma\delta+(\gamma^2+\delta^2)\alpha\beta}.
\end{eqnarray}
For the AKLT point, $\alpha=-\beta$ and $\delta$ (or $\gamma$)$=0$, $\gamma$ (or $\delta$)$=2\alpha$, $\Delta_{rung}=1$ and the R\'{e}nyi entropy indeed corresponds to the free energy of a critical model with central charge $c=1$\cite{PhysRevB.90.075151,PhysRevB.90.085137,Santos2015} as expected.
%%%%%%%%%%%%%%%%%%%%%%%
\subsubsection{Tracing one leg}\label{sec:leg}
%%%%%%%%%%%%%%%%%%%%%%%%%%%

\begin{figure}[h!]
	\centering
		\includegraphics[width=0.9\linewidth]{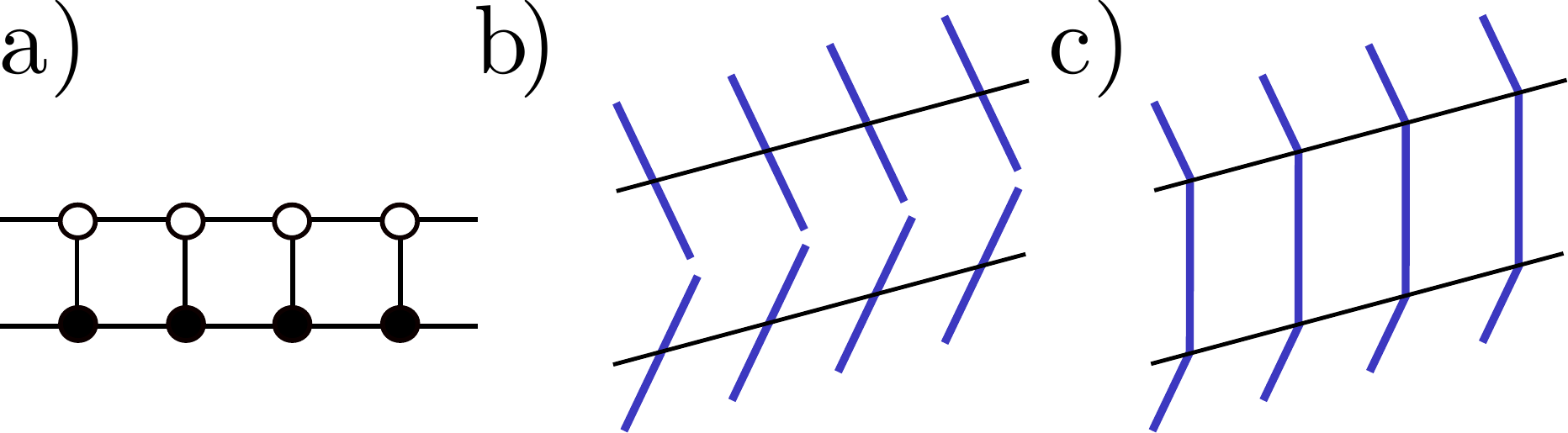}
	\caption{(Color online) a) Each leg of the ladder belongs to a different subset, represented here by white 
	and black dots. b) Tracing out the sites belonging to $A^c$. c) The resulting transfer matrix corresponds to $\rho_A$.}\label{fig:leg_partition}
\end{figure}

Another partition that we consider is the partition between the legs of the ladder, which give us information about the
entanglement between them. Critical entanglement spectrum has been shown to appear between the legs of spin-ladders 
\cite{PhysRevLett.105.077202,FurukawaKim:prb11,Lauchli:prb12,2012JSMTE..11..021S,Fradkin_Ladder,PhysRevB.88.245137}. Tracing out the spin states in $A^c$ (which is defined by the black dots in Fig. (\ref{fig:leg_partition}a)), we obtain $\rho_A$ as depicted in Fig. (\ref{fig:leg_partition}c). Stacking $n$ of them to form ${\rm tr}\rho_A^n$, we obtain again an eight vertex partition function, but with the Boltzmann weights
\begin{eqnarray}\nonumber\label{boltz_weights}
\omega_1=M_{-+}^{[+,-]}̣̣&=&\omega_2= M_{+-}^{[-,+]} =\gamma,\\\nonumber
\omega_3=M_{--}^{[+,+]}̣̣&=&\omega_4= M_{++}^{[-,-]}̣̣ =\beta,\\
\omega_5=M_{+-}^{[+,-]}̣̣&=&\omega_6= M_{-+}^{[-,+]}̣̣ =\delta,\\\nonumber
\omega_7=M_{++}^{[+,+]}̣̣&=&\omega_8= M_{--}^{[-,-]}̣̣ =\alpha.
\end{eqnarray}
 where the $\omega_i$ correspond to the different arrow configurations that define the eight vertex model, as in Fig.
(\ref{fig:TM_rungs}ii) rotated counterclockwise 45 degrees. 

Defining the matrix $(\tilde{T}_{leg})_{bd}̣^{ca}=M_{cd}^{[a,b]}$, we have
\begin{equation}\label{TM_leg}
 \tilde{T}_{leg}= 
\begin{pmatrix}
  \alpha & 0 & 0 & \delta\\
  0 & \gamma & \beta & 0 \\
 0  & \beta & \gamma  & 0 \\
 \delta & 0 & 0  & \alpha
 \end{pmatrix},
\end{equation}
 in this case 
\begin{equation}\label{Delta_leg}
 \Delta_{leg}=\frac{\alpha^2+\beta^2-\gamma^2-\delta^2}{2(\alpha\beta+\gamma\delta)},\quad
 \Gamma_{leg}=\frac{\alpha\beta-\gamma\delta}{\alpha\beta+\gamma\delta}\\.
\end{equation}
For the AKLT point, $\Delta_{leg}=1$ signaling that the entanglement between the legs of the ladder in the AKLT-like ground
state is also described by a CFT with central charge $c=1$. This provides a new and additional analytical proof of the numerical results of Ref.~\cite{PhysRevLett.105.077202}.

%%%%%%%%%%%%
\subsubsection{Zig-zag tracing}\label{sec:zig}
%%%%%%%%%%%%
 
\begin{figure}[h!]
	\centering
		\includegraphics[width=0.9\linewidth]{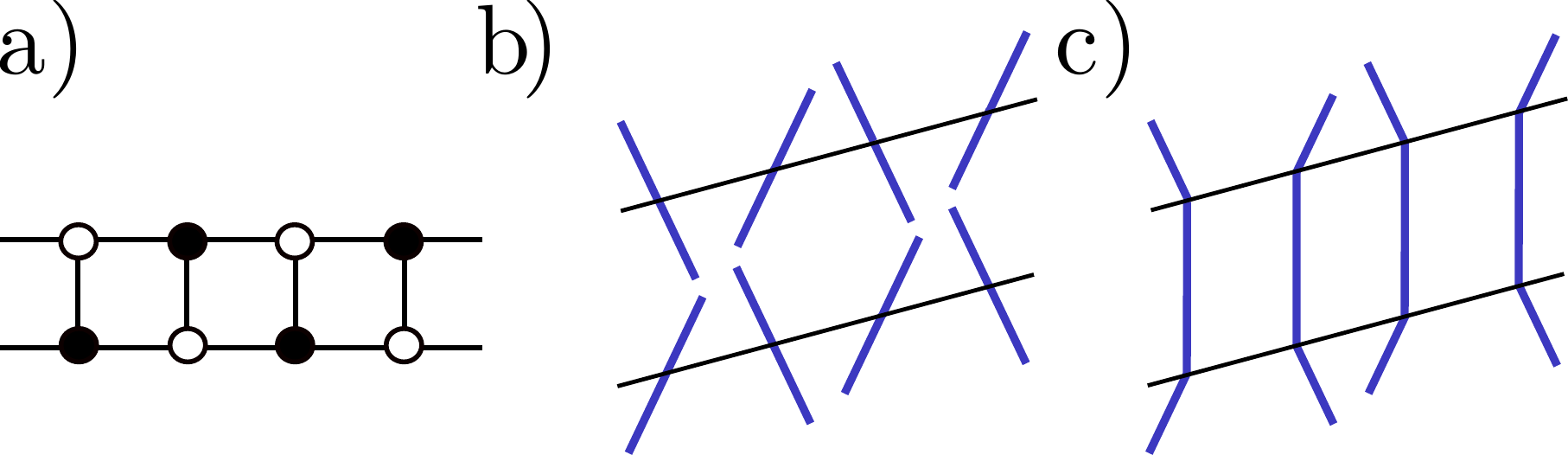}
	\caption{a) Sites in white belong to one subset $A$ and sites in black belong to $A^c$.
	b) Tracing of sites in $A^c$. In order to trace the sites in $A^c$, we contract the spin states in those sites. c) The operator that results from the tracing of the sites in $A^c$ corresponds to $\rho_A$.}\label{fig:zz_partition}
\end{figure}

Finally, for the MPS defined by Eq. (\ref{MPS_spin_half}), we analyze the entanglement across an alternating partition,
depicted in Fig. (\ref{fig:zz_partition}a). To our knowledge, this is the first time such a partition has been considered in spin ladder. The main difference from the previous partition is that the Boltzmann weights of the
resulting classical partition function are flipped in consecutive columns, so obtaining the corresponding eight vertex model is
not that straightforward as before. 
 
The Boltzmann weights in this case become the same as in (\ref{boltz_weights}) in one row, while in the next row we have to 
interchange $\delta$ and $\gamma$. Let the column transfer matrix be $Q(\alpha,\beta,\gamma,\delta)$. For the zig-zag partition, the R\'{e}nyi entropy of a ladder of length, $N$, becomes

\begin{equation}\label{Renyi_zz}
 \mathcal{S}_n=\frac{\ln\left({\rm tr}(Q_+Q_-)^{N/2}\right)}{1-n}.
\end{equation}

\noindent Here the matrix $Q_+=Q(\alpha,\beta,\gamma,\delta)$ and $Q_-=Q(\alpha,\beta,\delta,\gamma)$. In the computation of $S_n$ 
The matrix $Q$ has dimension $2^n\times 2^n$. From (\ref{Renyi_zz}) it is clear that the role of $\gamma$ and 
$\delta$ is interchanged from column to column. This modification does not change the value of the partition function, 
as the eight vertex model transfer matrix that construct the lattice {\it column by column} has eigenvectors that depend only 
on the combinations \cite{baxter2013exactly}

\begin{equation}
 \Delta_{leg}=\frac{\alpha^2+\beta^2-\gamma^2-\delta^2}{2(\alpha\beta+\gamma\delta)} \quad\mbox{and} \quad \Gamma=\frac{\alpha\beta-\gamma\delta}{\alpha\beta+\gamma\delta}.
\end{equation}

\noindent This indicates that the column transfer matrix $Q_+$ commutes with $Q_-$. The largest eigenvalue of $Q$ does 
depend on symmetric combinations of $\gamma$ and $\delta$, so in the limit of $N\gg 1$ we have

\begin{equation}\label{Delta_zigzag}
 \Delta_{leg}=\Delta_{zig-zag}.
\end{equation}

%%%%%%%%%%%%%%%%%%%%%%%%
\subsection{Bulk Entanglement Hamiltonian}
%%%%%%%%%%%%%%%%%%%%%%%%
In the previous section, we have shown how the R\'{e}nyi entropy $\mathcal{S}_n$ for three different partitions of the ground state (described by Eq. (\ref{MPS_spin_half_sol})) is equivalent to an eight vertex model partition function. We now show this connection allows us to have access
to the bulk entanglement Hamiltonian, $H_e$. $H_e$ is defined as the logarithm of $\rho_A$,

\begin{equation}
 \rho_A^n=\frac{e^{-n H_e}}{Z_0^n},
\end{equation}

\noindent where $Z_0$ is the norm of the ground state, in our case $Z_0=(\alpha^2+\beta^2+\gamma^2+\delta^2)^N$, where $N$
is the length of the ladder. We set this normalization to one.

As discussed previously, the R\'{e}nyi entropy is build up by stacking $\rho_A$ $n$-times
vertically. The operator $\rho_A$ is in turn built from the one dimensional transfer matrix $\tilde{T}$, where $\tilde{T}$ is defined accordingly
for each partition. As $\rho_A$ builds up the lattice, it is equivalent to the transfer matrix of the 8-vertex model. The 
logarithmic derivative of the 8-vertex model transfer matrix corresponds to the XYZ Hamiltonian {(except at certain parameter points, which we discuss later)}  \cite{baxter2013exactly}. But the logarithmic derivative
of the transfer matrix, and consequently $\rho_A$, is nothing more than $H_e$. This establish the 
correspondence $H_e^p=H^p_{XYZ}(\alpha,\beta,\gamma,\delta)$, where $p$ represents the different partitions. This is in agreement with earlier work on the. More explicitly
in terms of the Pauli matrices $\sigma_x=\sigma^++\sigma^-$ and $\sigma_y=-i(\sigma^+-\sigma^-)$ we have

\begin{equation}\label{ent_Ham_XYZ}
 H_e^{p}=-J_p\sum_{i=1}^N \sigma^x_{i}\sigma_{i+1}^x+\Delta_p\sigma^y_{i}\sigma_{i+1}^y+\Gamma_p\sigma^z_{i}\sigma_{i+1}^z,
\end{equation}

\noindent with $\Delta_{rung,leg,zig-zag}$ defined in Eqs. (\ref{Delta_rung},\ref{Delta_leg},\ref{Delta_zigzag}). $J_p$
and $\Gamma_p$ are 

\begin{eqnarray}
 J_{rung}&=&{\rm sgn}((\alpha^2+\beta^2)\gamma\delta+(\gamma^2+\delta^2)\alpha\beta),\\
 \Gamma_{rung}&=&\frac{(\alpha^2+\beta^2)\gamma\delta-(\gamma^2+\delta^2)\alpha\beta}{(\alpha^2+\beta^2)\gamma\delta+(\gamma^2+\delta^2)\alpha\beta},\\
 J_{leg}&=&J_{zig-zag}={\rm sgn}(\alpha\beta+\gamma\delta),\\
  \Gamma_{leg}&=&\Gamma_{zig-zag}=\frac{\alpha\beta-\gamma\delta}{\alpha\beta+\gamma\delta}.
\end{eqnarray}

\noindent As pointed out in \cite{PhysRevB.90.085137}, $H^p_e$ inherits the same symmetries as the original MPS. At the isotropic point, $\delta$ $(\gamma)=0$ and $\alpha-\beta=\gamma$ $(\delta)$, Eq. (\ref{ent_Ham_XYZ}) has full rotational symmetry
and corresponds to the spin 1/2 Heisenberg model.
In general, depending on the parameters, $H_{XYZ}$ can be ferro or antiferromagnetic. 
In both cases the spectrum of $H_{XYZ}$ has a degenerate ground state (in the thermodynamic limit) or 
a unique ground state with a gapless excitations. We see explicitly that for this model, time reversal
symmetry or $\mathbb{Z}_2\times\mathbb{Z}_2$ imply a BES either gapless or with degenerate ground state levels.

%%%%%%%%%%%%%%%%%%
\subsection{Critical phases of bulk entanglement}
%%%%%%%%%%%%%%%%%%
In this section, we analyze the relation between the critical phases of $H_e$ for the three types of 
partitions discussed so far. For the MPS we consider, the critical phases of $H_e$ for rung, leg and zig-zag partitions coincide. After some straightforward algebra, it is possible to show that $\Delta_{rung}=\pm 1$ implies 
$\Delta_{leg}=\pm 1$ {\it and} that $\Delta_{leg}=\pm 1$ implies $\Delta_{rung}=\pm 1$. So the bulk entanglement becomes
critical in all partitions considered above simultaneously. We write this relation as
\begin{equation}
 \Delta_{rung}=\pm 1 \leftrightarrow \Delta_{leg}=\pm 1.
\end{equation}
Similarly, the relation $\Gamma_{rung}=\pm 1 \leftrightarrow \Gamma_{leg}=\mp 1$ also holds.

We can understand why $H_e$ becomes critical when $\Delta_{leg}=\pm 1$ by considering the transfer matrix (Eq. (\ref{TM_rung})). The eigenvalues of the transfer matrix are

\begin{eqnarray}
 \lambda_1&=&\alpha^2+\beta^2+\gamma^2+\delta^2,\\
 \lambda_2&=&\alpha^2+\beta^2-\gamma^2-\delta^2,\\
 \lambda_3&=&2(\alpha\beta+\gamma\delta),\\
 \lambda_4&=&2(\alpha\beta-\gamma\delta).
\end{eqnarray}

\noindent The condition $\Delta_{leg}=\pm 1$ (and consequently a critical $H_e$ for all three partitions) is simply the 
condition that $\lambda_2=\pm \lambda_3$. The correlation functions and in particular the correlation lengths in the ladder 
are precisely controlled by the ratios $\frac{\lambda_2}{\lambda_1},\frac{\lambda_3}{\lambda_1}$ (Eqs. (\ref{Corr_lenght_x}-\ref{Corr_lenght_z})), 
so we conclude that whenever two correlation lengths coincide, the entanglement in the bulk becomes critical
for this MPS.

We now discuss the BES of the NT point in detail. If $\alpha\beta+\gamma\delta=0$, which is the case at the NT point, the 
connection between $H_e$ and the XYZ model breaks down. As mentioned earlier, the NT wave function is simply one possible dimer 
covering of a spin ladder (corresponding to the red bonds in Fig. (\ref{fig:Spin_S_ladder}), without the symmetrization in each 
rung). Thus, the BES is \emph{flat} and the bulk entanglement entropy for the alternating rung partition (Sec. (\ref{sec:bes})) 
is simply a measure of how many dimers are present across the partition. This also implies that the zig-zag partition and the leg 
partition are not longer equivalent. For the leg partition (Sec. (\ref{sec:leg}), all the singlets belong to the two 
complementary sets $A$ and $A^c$ making
the entanglement entropy maximal between these sets. For the zig-zag cut (Sec. (\ref{sec:zig}), the opposite is true. The singlets belong to either $A$
or $A^c$ so the ground state wave function is separable and the entanglement entropy is zero. $\rho_A$ is 
proportional to the identity for the leg partition, and a projector onto the ground state in the zig-zag partition.
This indicates that for $u\rightarrow \pm1$, $H^{leg}_{XYZ}\rightarrow 0$ while $H^{zig-zag}_{XYZ}\rightarrow \infty$.
This result is consistent with the results in Ref.~\cite{Lundgren} at the NT point, up to an unitary transformation in every other 
rung of the groundstate that permutes the spin states. This unitary transformation translates the MPS representation of our work into the 
ground state studied in Ref.~\cite{Lundgren} at the NT point.
%%%%%%%%%%%%%%%%%%
%\subsection{Structure of bulk entanglement spectrum}
%%%%%%%%%%%%%%%%%%%
%{\color{red}\sout{ As pointed out in \cite{PhysRevB.90.085137}, the entanglement Hamiltonian (\ref{ent_Ham_XYZ}) inherits the same symmetries as the original MPS.
%At the isotropic point $\delta$ $(\gamma)=0$ and $\alpha-\beta=\gamma$ $(\delta)$ the BE Hamiltonian has full rotational symmetry
%and corresponds to the spin 1/2 Heisenberg model.
%In general, depending on the parameters, $H_{XYZ}$ can be ferro or antiferromagnetic. 
%In both cases the spectrum of $H_{XYZ}$ has a degenerate ground state (in the thermodynamic limit) or 
%a unique ground state with a gapless excitations. We see explicitly that for this model, time reversal
%symmetry or $\mathbb{Z}_2\times\mathbb{Z}_2$ imply a BES either gapless or with degenerate ground state levels.
%.} (I moved this section to other parts where it fits better)}

%%%%%%%%%%%%%%%%%%%%%
\section{Spin-$S/2$ ladder state and connection to spin $S$ AKLT}\label{sec:spinSladder}
%%%%%%%%%%%%%%%%%%%%
 In this section, we will argue that, given that the AKLT state is in the Haldane phase, 
the structure of the BES is a consequence of the Lieb-Schultz-Mattis theorem 
\cite{Lieb1961,Affleck1986}. We note, unlike the spin-1 AKLT state, the spin-$S$ AKLT state can be connected 
 to a topologically trivial state for even spin $S$ \cite{PhysRevB.85.075125,PhysRevB.87.235106}. As the spin-$S$ AKLT state can be 
 realized in a spin $S/2$ ladder, we would like to see how this difference between even and odd integer spins is reflected in the 
 BES of spin ladders with $S>1$. To this end, we extend the previous discussion to spin $S/2$ particles on each leg of the ladder. We consider partitions where groups of rungs are traced out. When specialized to the AKLT state, these partitions reduces to tracing every other or every
 two other sites.
 
We consider the following MPS, which is generalization of the AKLT state (in spin ladders),
\begin{equation}\label{gen_MPS}
 (A_i(S,\vec{\gamma}))_{ab}=\sum_{j,m}\gamma_j\ClebschG{\frac{S}{2}}{a}{\frac{S}{2}}{-b}{j}{m}(-1)^{S/2+b}\ket{j,m}_i,
\end{equation}
where $\ClebschG{\frac{S}{2}}{a}{\frac{S}{2}}{b}{j}{m}\equiv\langle S/2,a;S/2,b|j,m\rangle$ are the Clebsch-Gordan 
coefficients for two spin S/2 particles and $\gamma_j$ is set of real parameters. The many-body ground state is $\ket{\psi(S,\vec{\gamma})_0}=\mathrm{tr}(A_1(\vec{\gamma})A_2(\vec{\gamma})\cdots A_N(\vec{\gamma}))$.
This MPS reproduces the spin $S$ AKLT ground state for $\gamma_j=\delta_{jS}$ the Kronecker delta \cite{Xu2008} (see
App. \ref{app:HigherSVBS}). This corresponds to the generalization of having only triplets in the spin-half ladder case.

Considering a bipartition where we trace every other rung (Fig. \ref{fig:rung_partition}a), the transfer matrix is
\begin{align}\label{transfer_m}
\tilde{T}_{ab}^{cd}=\sum_{j,m}\gamma_j^2\ClebschG{\frac{S}{2}}{a}{\frac{S}{2}}{-b}{j}{m}\ClebschG{\frac{S}{2}}{c}{\frac{S}{2}}{-d}{j}{m}(-1)^{S+b+d}.
\end{align}
The right-left eigenvalues of this operator are given in terms of the $6j$ symbols \cite{Santosqdef} (see App. (\ref{appendixB}))
\begin{equation}\label{eig_TM}
 \lambda_j^{\{\gamma\}}=(-1)^j\sum_{J=0}^S\gamma^2_J(-1)^{S+J}(2J+1)\SixJun{J}{j}{\frac{S}{2}}{\frac{S}{2}}{\frac{S}{2}}{\frac{S}{2}}.
\end{equation}
Note that each $\lambda_j^{\{\gamma\}}$ eigenvalue is $2j+1$ degenerate due to $SU(2)$ symmetry. 
%If $\gamma_j=1$ for all $j$, the transfer matrix takes a special 
%form $\tilde{T}_{ab}^{cd}=\delta_{a}^c\delta_{b}^d$ {\color{red}Do we use this property? Is it true for the ALKT state?}. 
Using the properties of the Clebsch-Gordan coefficients, the transfer matrix can be brought to the spectral form \cite{Santosqdef}
\begin{eqnarray}
 \tilde{T}_{ab}^{cd}&=&\sum_{j,m}\lambda_j^{\{\gamma\}}\ClebschG{\frac{S}{2}}{-a}{\frac{S}{2}}{c}{j}{m}\ClebschG{\frac{S}{2}}{-b}{\frac{S}{2}}{d}{j}{m}(-1)^{S+a+b}.~
\end{eqnarray}
For future reference, we write also the transfer matrix a general bipartition, where we trace $k$ consecutive rungs,
\begin{equation}
 (\tilde{T}^k)_{ab}^{cd}=\sum_{j,m}R_j^{k,\{\gamma\}}\ClebschG{\frac{S}{2}}{a}{\frac{S}{2}}{-b}{j}{m}\ClebschG{\frac{S}{2}}{c}{\frac{S}{2}}{-d}{j}{m}(-1)^{S+b+d},
\end{equation}
with
\begin{equation}
 R_j^{k,\{\gamma\}}=\sum_{J=0}^S(\lambda_J^{\{\gamma\}})^k(2J+1)\SixJun{\frac{S}{2}}{\frac{S}{2}}{\frac{S}{2}}{\frac{S}{2}}{j}{J}(-1)^{j-J+S}.
\end{equation}
From the previous results, it follows that the transfer matrix can be written in the form
\begin{equation}
 (\tilde{T}^k)_{ab}^{cd}=(\lambda_0^{\{\gamma\}})^k\left[\frac{\delta_{a}^c\delta_b^d}{S+1}+\sum_{j=0}^SQ_j^{k,\{\gamma\}}U\hat{\Pi}_j^VU^\dagger\right],
\end{equation}
where
\begin{equation}\label{Qjs}
Q_j^{k,\{\gamma\}}=\sum_{J=1}^S\left(\frac{\lambda_J^{\{\gamma\}}}{\lambda_0^{\{\gamma\}}}\right)^k(2J+1)\SixJun{\frac{S}{2}}{\frac{S}{2}}{\frac{S}{2}}{\frac{S}{2}}{j}{J}(-1)^{j-J+S}.
\end{equation}
Here, $(U)_{aa',bb'}=(-1)^{S/2+b}\delta_{-bb'}\delta_{aa'}=(\mathbb{I}\otimes e^{-i\pi S^y})_{aa',bb'}$ with $S^y$ the spin-$S/2$ 
operator
(which acts in the auxiliary space of the bond indices for the MPS) and $\hat{\Pi}^V_J$ is the {\it vertical} projector operator onto the total spin $J\in[0,S]$ channel
\begin{equation}\label{prod}
\hat{\Pi}^V_{J}(\vec{S}_i\cdot\vec{S}_{i+1})=\prod_{j=0,j\neq J}^S\frac{S(S+2)+4\vec{S}_i\cdot\vec{S}_{i+1}-2j(j+1)}{2J(J+1)-2j(j+1)},
\end{equation}
with $i$ and $i+1$ corresponding to consecutive auxiliary spaces. Note that this operator is a projector acting from vertically from bottom to top.
Defining the ratio $r_j^{k,\{\gamma\}}=(\lambda_j^{\{\gamma\}}/\lambda_0^{\{\gamma\}})^k$
which defines the correlation length $\xi_j\{\gamma\}$ i.e. $r_j^{k,\{\gamma\}}\sim e^{-k/\xi_j\{\gamma\}}$, so  
$r_j^{k,\{\gamma\}}\ll 1$ for gapped states \footnote{Note that this approximation is increasingly good for bipartitions that contain many consecutive rungs}, we can approximate $\rho_A$ obtained from the transfer matrix (Eq. (\ref{transfer_m})) up to first order in $Q_j^{k,\{\gamma\}}$. In this approximation, $H_e$ takes the form
\begin{equation}\label{ent_ham_VBS_hor}
H_e=-(S+1)\sum_{i=1}^{N_{\rm sites}} \sum_{J=0}^S Q_J^{k,\{\gamma\}} \hat{\Pi}^V_{J}(\vec{S}_i\cdot(u\vec{S}u^\dagger)_{i+1}).
\end{equation}
By performing the unitary transformation $u=e^{-i\pi S_y}$ on every other site, 
we can simplify the previous expression using that $u_{i+1}\hat{\Pi}_J(\vec{S}_i\cdot (u\vec{S}u^\dagger)_{i+1}){u}^{\dagger}_{i,i+1}=\hat{\Pi}_J(\vec{S}_i\cdot\vec{S}_{i+1})$,
The effective Hamiltonian $H_e$ becomes
\begin{equation}\label{BEH_projectors}
H_e=-(S+1)\sum_{i=1}^{N_{\rm sites}}\sum_{j=0}^SQ_j^{k,\{\gamma\}} \hat{\Pi}^V_{j}(\vec{S}_i\cdot\vec{S}_{i+1}).
\end{equation}
Eq. (\ref{BEH_projectors}) corresponds to an effective Hamiltonian of a spin $S/2$ chain. At the VBS point ($\gamma_J=\delta_{JS}$), for the bipartition where we trace every other rung, $H_e$ is given by
\begin{equation}
 H_e^{(1)}=-\sum_{i=1}^{N_{\rm sites}}\frac{(S+1)^2}{2S+1}\hat{\Pi}^V_S(\vec{S}_i\cdot\vec{S}_{i+1}),
\end{equation}
which corresponds to a {\it ferromagnetic} {spin S/2 chain}. For the other bipartition, where we trace every other two rungs,
the BEH of the spin $S$ VBS becomes
\begin{equation}\label{BEH_antiferro}
 H_e^{(2)}=-(S+1)^3\sum_{i=1}^{N_{\rm sites}}\sum_{J=0}^S\SixJun{S}{\frac{S}{2}}{S}{\frac{S}{2}}{J}{\frac{S}{2}}^2\hat{\Pi}^V_J(\vec{S}_i\cdot\vec{S}_{i+1}),
\end{equation}
up to an constant. This BEH is {\it antiferromagnetic}. This can be seen clearly in the limit $S\gg1$ where
\begin{equation}
 \SixJun{S}{\frac{S}{2}}{S}{\frac{S}{2}}{J}{\frac{S}{2}}^2\approx f(S)\frac{}{}e^{\frac{-J^2}{2S}}\left[1+\mathcal{O}\left(\frac{1}{\sqrt{S}}\right)\right]
\end{equation}
with $f(S)$ given in Appendix \ref{app:6j}. This even odd effect as a result of the partition appears
due to the Neel-like order of VBS ground state.

The results for the generalized AKLT state can be summarized as follows. For $S$ an odd integer, $H_e$ corresponds to a 
ferromagnetic/antiferromagnetic Hamiltonian of a half integer spin, depending on the partition considered. In both cases the 
BES is gapless or possess a double degenerate ground state. For the antiferromagnetic case, this is proved by the 
Lieb-Mattis-Schultz theorem\cite{Lieb1961}. For $S$ an even integer, tracing every two other rungs of the ladder, the BEH is 
given by (\ref{BEH_antiferro}) which corresponds to an integer spin chain, with {\rm antiferromagnetic} coupling. Generically, 
following Haldane's analysis \cite{PhysRevLett.50.1153}, the BES (for even S) for this partition is expected to have an 
entanglement gap.
%%%%%%%%%%%%%%%%%%
\subsection{Degeneracy Preserving Deformations}
%%%%%%%%%%%%%%%%%%%%
In this section, we investigate deformations of the AKLT state. As we discussed in detail in for the $S=1/2$ ladder, when the eigenvalues of the transfer matrix coincide, $H_e$ can become critical. For the general spin case, we also expect
that extra level crossings in the transfer matrix could change the low lying properties of $H_e$.  Usually, to argue that two states belong to the same phase, it is enough to show that during the 
interpolation the largest eigenvalue of the transfer matrix remains unique, to avoid a quantum phase transition (or
more specifically, a divergent correlation length \cite{nakahara2012frontiers}). This condition is clearly not enough to assure that
the resulting transfer matrix faithfully represents the phase of the initial entanglement Hamiltonian. An extreme 
example corresponds to the interpolation between an arbitrary $N^2\times N^2$ transfer matrix $(T_0)_{ab}^{cd}$ 
(written in components, with $a,b,c,d = 1..N$)
 and the identity matrix $(T_1)_{ab}^{cd}=\delta_{a}^c\delta_{b}^d$. Assuming that $T_0$ has a unique largest eigenvalue 
$\lambda_{\rm max}>0$, we can deform its spectrum to match the spectrum of the identity matrix $T_1$. This amounts to deform 
$\lambda_{\rm max}\rightarrow N$ and the rest of the eigenvalues $\{\lambda\}\rightarrow 0$. In this deformation, the gap between 
the largest and the sub-leading eigenvalue never closes, so the two states are connected without a phase transition. 
However, the final transfer matrix generates a vanishing $H_e$, erasing all the properties
of the original $H_e$ associated with $T_0$ (an explicit example of a nontrivial deformation which shows a critical BEH 
is discussed in detail in the next section). In view of this, we require a stronger condition on the allowed deformations, namely: A deformation of the transfer matrix preserves the structure of $H_e$ if
does not generate extra degeneracies during the deformation. We denote such deformation as a
{\it Degeneracy Preserving} (DP) deformation.

\begin{figure}[h!]
	\centering
		\includegraphics[width=1\linewidth]{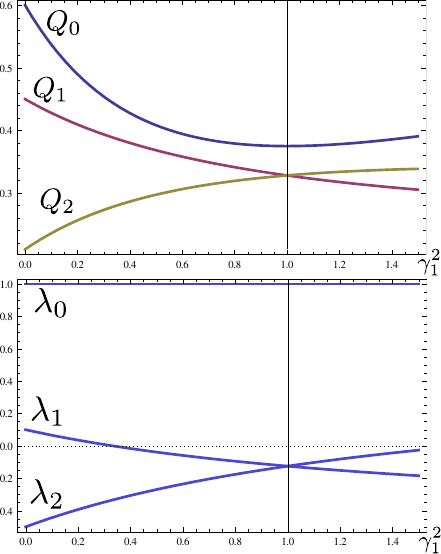}
	\caption{(Color online) (Top) For $S=2$ we set $\gamma_2=0$, $\gamma_0=0$ and vary $\gamma_1$. We plot the relative weights $Q_j$
	as we vary $\gamma_1^2$ between 0 and 1.5. At $\gamma_1=1$, the weights $Q_1$ and $Q_2$ coincide.
	(Bottom) Plot of the eigenvalues $|\lambda_j(\gamma_1)/\lambda_0(\gamma_1)|$ of the transfer matrix,
	for $\gamma_1^2$ between 0 and 1.5. The VBS point corresponds to $\gamma_1=0$. As a function of $\gamma_1$
	the deformation is DP for $\gamma_1<1$. At $\gamma_1=1$, the second and third eigenvalues of $\hat{T}$ coincide and
	the BES becomes critical.}
	\label{fig:Interpol}
\end{figure}

We now illustrate this point for $S=2$. We construct a non-DP deformation that explicitly changes the properties
of the BES. To do this, we deform the spectrum of the one dimensional transfer matrix $\hat{T}$, showing that a change in the 
degeneracy of $\hat{T}$ implies a change in the BES. For $S=2$, taking $\gamma_{2}=1$, $\gamma_{0}=0$ and leaving $\gamma_1$ as a free parameter, we plot the relative weights $Q_j$, Eq. (\ref{Qjs}), as a function of $\gamma_1$. We also plot the eigenvalues $\lambda_j$, Eq. (\ref{eig_TM}),
as a function of $\gamma_1$. As shown in Fig. \ref{fig:Interpol} the eigenvalues of the transfer matrix coincide for
$\gamma_1=1$. At this point, the BEH can be written as
\begin{eqnarray}\nonumber\label{special2}
 H_e^{(2)}&=&-3\sum_{i=1}^{N_{\rm sites}}\frac{3}{8}\hat{\Pi}_0^V(i)+\frac{21}{64}(\hat{\Pi}_1^V(i)+\hat{\Pi}_2^V(i))\\
 &=&-3\sum_{i=1}^{N_{\rm sites}}\frac{3}{64}\hat{\Pi}_0^V(i)+\frac{21}{64}.
\end{eqnarray}
Here, we have used the completeness relation, $\sum_{J=0}^S\hat{\Pi}_J^V=\mathbb{I}$. The Hamiltonian, Eq. (\ref{special2}), consists
of a sum of projectors onto the singlet state between two consecutive sites. The ground state of this Hamiltonian 
is doubly degenerate. It breaks spontaneously translation symmetry and consists of one of the two dimer coverings of 
the one dimensional lattice \cite{Affleck1990}. We thus see explicitly that a change in the degeneracy of the MPS transfer matrix drastically changes the low lying BES.

\subsection{BES for Spin $S$}
 We find that for the case of tracing every other site, $H_e^{(1)}$ is {\it ferromagnetic}, while tracing every two other sites,
 $H_e^{(2)}$ is {\it antiferromagnetic}. This remains true for any deformation that does not change the degeneracy of the transfer
 matrix. In particular is valid for small deformations around the VBS point $\gamma_i\ll1$ with $i=0,1,...,S-1$. The general entanglement
 Hamiltonian in that case is given by (\ref{BEH_projectors}).

For $S$ odd, both the ferromagnetic and the antiferromagnetic Heisenberg models have a gapless spectrum 
\cite{PhysRevLett.50.1153,Lieb1961,Affleck1986}, so the BES in this case is gapless. For $S$ even, the ferromagnetic Hamiltonian $H_e^{(1)}$ is gapless, while the antiferromagnetic Hamiltonian $H_e^{(2)}$ is gapped. This indicates that the gapless BES is 
gapped for $S$ even. This constitutes one of the main results of this article, namely {\it the BES of the VBS ground state realized in a spin $S/2$ ladder is gapped for $S$ even and gapless (or double degenerate) for $S$ odd}. The ground state of the higher spin AKLT is a representative of the Haldane phase for arbitrary spin $S$, implying that the BEH of the Haldane phase with $S$ odd is gapless. For $S$ even, the BEH is generically not critical.

%%%%%%%%%%%%%
\section{Exact deformation}\label{sec:NON_UNI}
%%%%%%%%%%%%%

In this section we discuss the non-universality of the BES under non-DP deformations. We now give an example of how non-DP deformations from the VBS point create extra degeneracies in the transfer matrix along the interpolation that dramatically change the low lying BES. This is done by introducing a non DP-deformation which does not induce a diverging correlation
length for any value of the interpolation parameter. The resulting state is shown to have a critical BES which can be extracted exactly, signaling that criticality of the BEH does not appear only in states with topological order. If a deformation is not of DP type as  defined in the previous section, the BEH can undergo a phase transition while the physical state does not. In other words, the BES does not capture the universal physics present at the phase transition. The non-universality of the ES of a real-space bipartition was first seen numerically in Ref.~\cite{PhysRevLett.104.180502} and later analytically explored in Ref.~\cite{Chandran2014}. 

\begin{figure}[th!]
	\centering
		\includegraphics[width=0.85\linewidth]{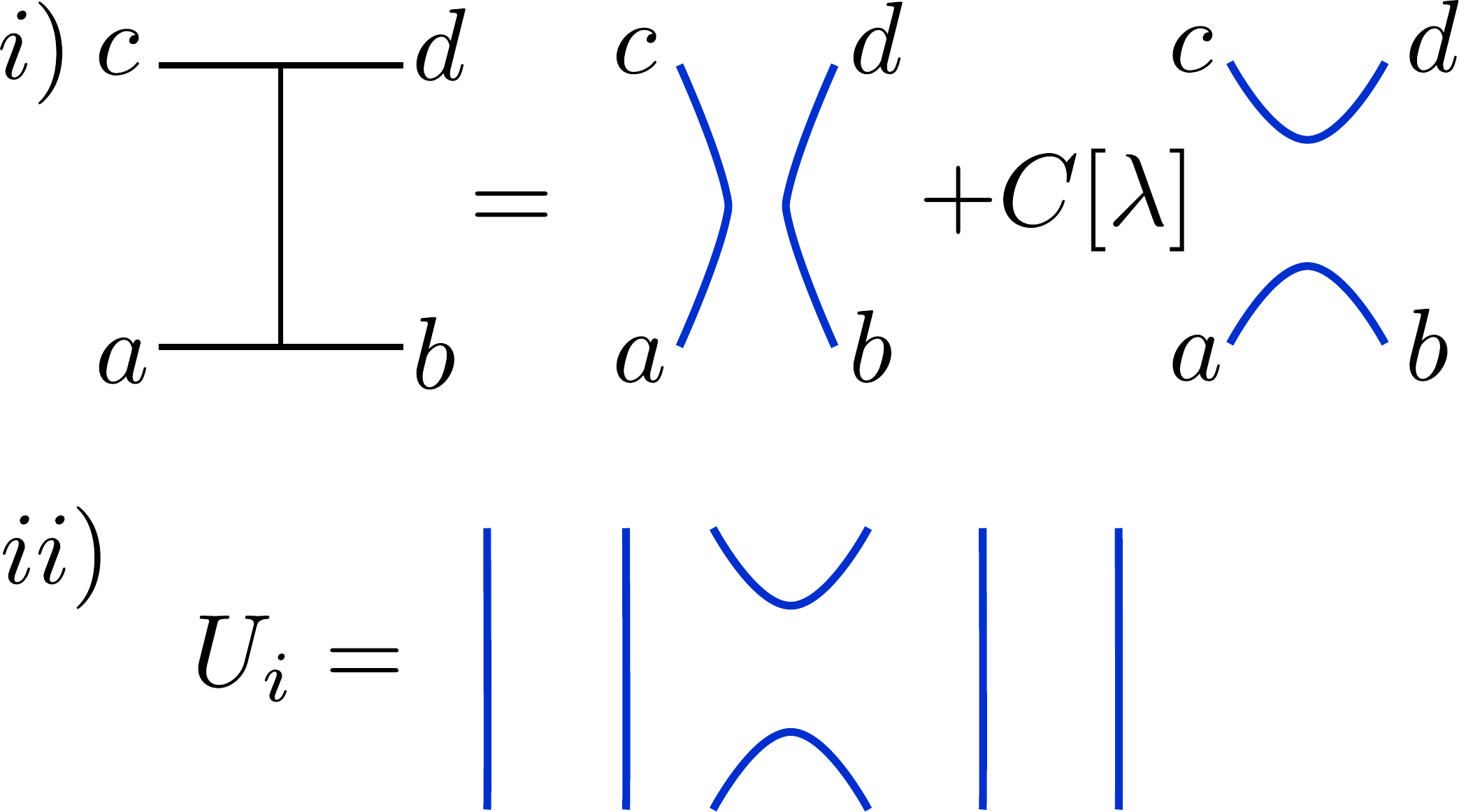}
	\caption{(i) Diagrammatic representations of transfer matrix, each line represents a $\delta$-function between
	the indices in the corresponding sites (ii) Diagrammatic definition of operator $U_i$. This operator acts 
	nontrivially just in sites $i$ and $i+1$.}\label{fig:TMS}
\end{figure}

\begin{figure}[th]
		 \includegraphics[width=1\linewidth]{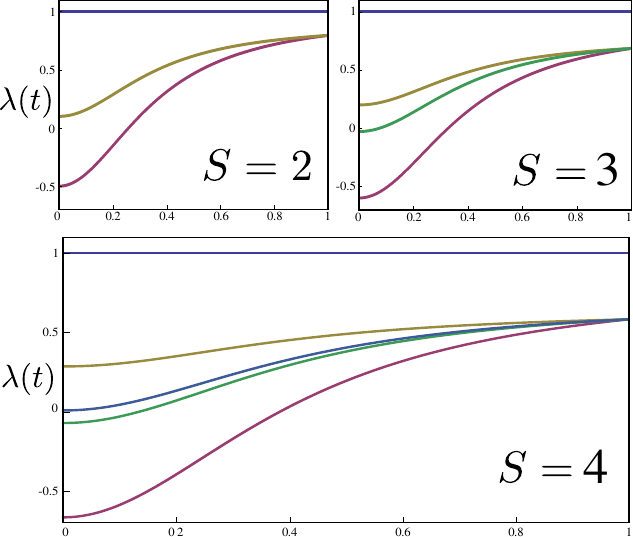}
	\caption{Transfer matrix $\tilde{T}_{ex}(t)$ spectrum for $S=2$ (upper left) $S=3$ (upper right)
	and $S=4$ (lower panel) as a function of the interpolation parameter $t$. Between the AKLT point $t=0$ and the exact mapping point $t=1$, all the subleading eigenvalues remain
	smaller than one, so the ground state correlations decay exponentially and no phase transition is observed.}\label{fig:INT}
\end{figure}

We explore this scenario by finding the exact R\'{e}nyi entropy
of a state connected to the spin $S$ VBS by the deformation
\begin{equation}\label{interpolation_exact}
 (G(t)_i)_{ab}=(1-t)(A_i(S,\delta_{SJ}))_{ab}+t(G_i)_{ab}
\end{equation}
where $A_i(S,\delta_{SJ})$ is the AKLT MPS and $G_i$ is given by
\begin{eqnarray}\label{def_gS}\nonumber
 ({G}_i)_{ab}&=&\sum_{J,m}{\ClebschG{\frac{S}{2}}{a}{\frac{S}{2}}{-b}{J}{m}}(1+\lambda\delta_{J0}\delta_{m0})(-1)^{S/2+b}\ket{J,m}_i,\\
 &=&(-1)^{S/2+b}|a,-b\rangle+\frac{\lambda(-1)^{S}}{\sqrt{S+1}}\delta_{ab}|s\rangle,
\end{eqnarray} 
where $|a,b\rangle$ is a compact notation for $\left|\frac{S}{2},a\right\rangle_{i,1}\otimes\left|\frac{S}{2},b\right\rangle_{i,2}$
and $|s\rangle$ is the singlet state.
The transfer matrix,$\tilde{T}_{ex}$, of this state at $t=1$ is 
\begin{eqnarray}\label{def_Tgs}
 (\tilde{T}_{ex})_{ab}^{cd}&=&\delta_{a}^c\delta_{b}^{d}+\frac{|\lambda+1|^2-1}{S+1}\delta_{ab}\delta^{cd},\\
 &=&\delta_{a}^{c}\delta_{b}^{d}+C[\lambda]\delta_{ab}\delta^{cd}.
 \end{eqnarray} 
 Upper and lower indices are used to distinguish the different spaces involved, but as tensors $\delta_{a}^{b}=\delta_{ab}=\delta^{ab}=1$ if $a=b$ and zero otherwise. The transfer matrix operator $\tilde{T}_{ex}$ can be 
easily interpreted diagrammatically (see Fig (\ref{fig:TMS}i)). 
 In general, the transfer matrix of $G_i(t)$ has eigenvalues, $\lambda_j(t)$, which are given by
 Eq. (\ref{eig_TM}) with $\gamma_J(t)=(1-t)\delta_{JS}+t(1+\lambda\delta_{J0}\delta_{m0})$. These eigenvalues are plotted
 along the interpolation $t\in[0,1]$ for $S=2,3$ and $4$ in Fig.~\ref{fig:INT}.

To analyze the R\'{e}nyi entropy of Eq. (\ref{interpolation_exact}), it is convenient to define the operator
\begin{equation}
 (U_i)_{a_1\dots a_N}^{b_1\dots b_N}=\left(\prod_{j\neq i,i+1} \delta_{a_j}^{b_j}\right)\delta_{a_ia_{i+1}}\delta^{b_ib_{i+1}}.
\end{equation}
$U_i$ is graphically represented in Fig. (\ref{fig:TMS}ii) and satisfies the Temperley-Lieb algebra \cite{Temperley1971}
\begin{eqnarray}
 U_i^2=(S+1)U_i,\\
 U_iU_{i\pm 1}U_i=U_i,\\
 U_iU_j=U_jU_i, \quad |i-j|>1.
\end{eqnarray}

\begin{figure}[tb]
	\centering
		\includegraphics[width=0.85\linewidth]{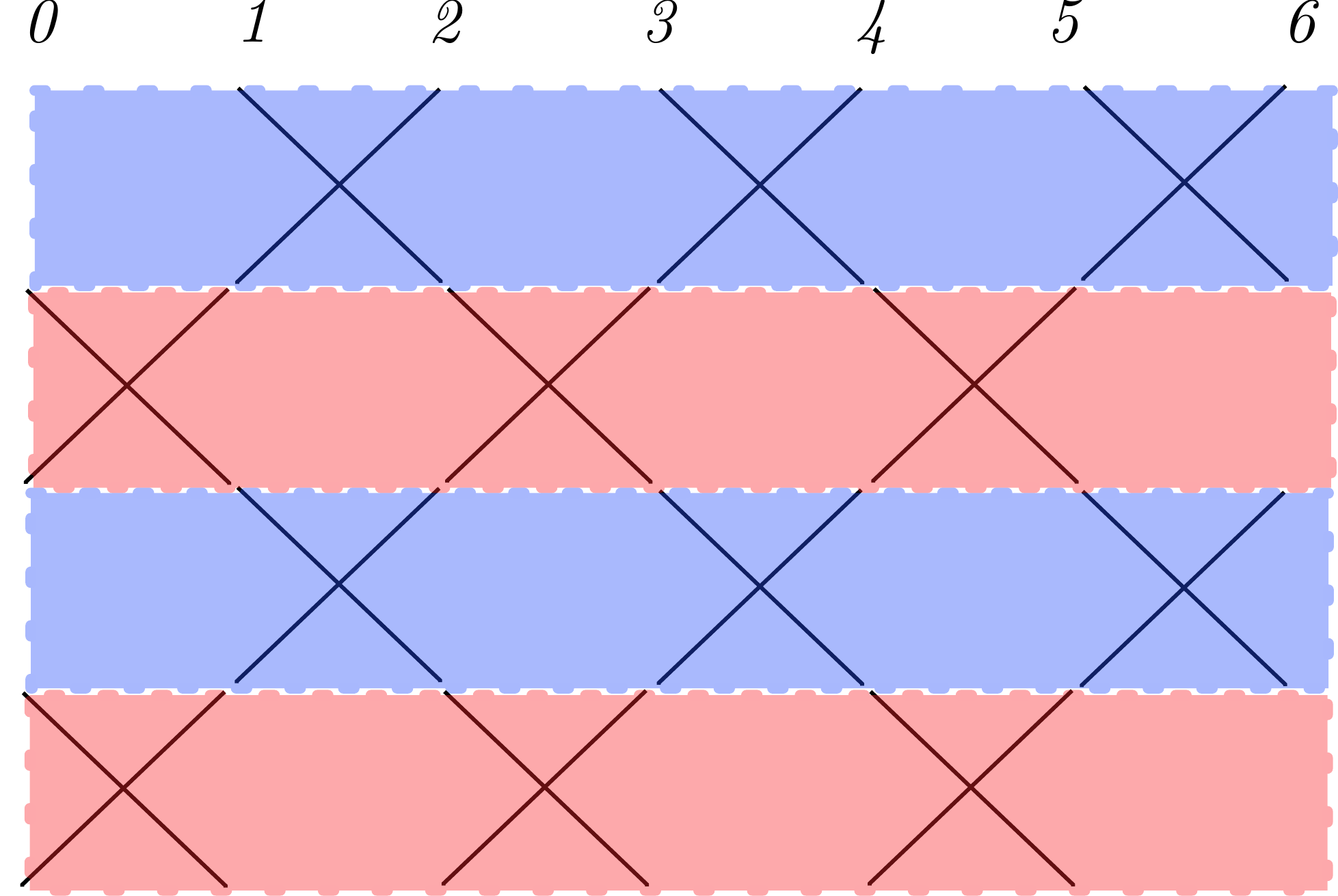}
	\caption{(Color online) Row by row construction of $\rho_A$. Each blue row corresponds to
	one application of the operator $V$. Each red row corresponds to one application of $W$.}\label{fig:Trans_Potts}
\end{figure}

\noindent Defining the identity operator $I$ that acts trivially on all the sites, we can write $\rho_A$ 
for this state as (see Fig. (\ref{fig:Trans_Potts}))
\begin{equation}
 \rho_A=\prod_{i=0}^{\frac{N}{2}-1}(I+C[\lambda] U_{2i+1})\prod_{i=1}^{\frac{N}{2}}(I+C[\lambda] U_{2i}),
\end{equation}
where we have used that the number of sites $N$ is even and periodic boundary conditions imply $N+1=1$. 
Using a different representation of the same Temperley Lieb (TL) generators, we can relate $\rho_A$ to 
the transfer matrix of the $(S+1)^2-$state Potts model \cite{Temperley1971,baxter2013exactly,Santos2015} as follows. 
The TL generators in a new representation $R$ can be written explicitly as
\begin{eqnarray}
 (U_{2i-1}^{R})_{\sigma}^{\sigma'}&=&\frac{1}{S+1}\left(\prod_{j=1\neq i}^{N/2}\delta_{\sigma_j}^{\sigma'_j}\right)\eta_{\sigma_i}^{\sigma_i'}\\
 (U_{2i}^{R})_{\sigma}^{\sigma'}&=&(S+1)\delta_{\sigma_i,\sigma_{i+1}}\prod_{j=1}^{N/2}\delta_{\sigma_j}^{\sigma'_j},
\end{eqnarray}
where $\eta_{\sigma}^{\sigma'}$ is an $(S+1)^2\times(S+1)^2$ matrix with all its entries equal to 1. Writing down the matrices $V=\prod_{i=0}^{\frac{N}{2}-1}(I+C[\lambda] U_{2i+1}^R)$ and $W=\prod_{i=1}^{N/2}(I+C[\lambda] U_{2i}^R)$, $\rho_{A,R}$ in this representation is simply $\rho_{A,R}=VW$. Now, using the TL algebra it is possible to rewrite the matrices $U$ and $V$ as \cite{Temperley1971,baxter2013exactly}
\begin{eqnarray}\nonumber
 V_{\sigma,\sigma'}&=&\exp\left(\sum_{i=0}^{\frac{N}{2}-1}K_2\delta_{\sigma_j\sigma_j'}\right),\\
 W_{\sigma,\sigma'}&=&\exp\left(\sum_{i=1}^{\frac{N}{2}}K_1\delta_{\sigma_i,\sigma_{i+1}}\right)\prod_{i=1}^{N/2}\delta_{\sigma_i,\sigma_i'},
 \end{eqnarray}
where $K_1=\ln((|\lambda+1|^2)$ $K_2=\ln\left(1+\frac{1}{|\lambda+1|^2-1}\right)$. The product of 
these matrices correspond to the transfer matrix of the two dimensional classical Potts model. The partition function, which 
given by ${\rm tr} \rho_{A,R}^\alpha$, only depends on the TL algebra \cite{Temperley1971}, so the change in representation 
establishes the mapping between the R\'{e}nyi entropy for the bulk entanglement and the partition function of the Potts model 
at criticality. 

The phase transition of the $q-$state Potts is of second order for $q\leq4$ and first order for $q>4$ \cite{Temperley1971}.
Based on this result, we conclude that for $S>2$ the R\'{e}nyi entropy, $\mathcal{S}_{n}$, of the 
alternating bipartition of Eq. (\ref{def_gS}) is non-analytic in the limit of $n\rightarrow \infty$, signaling a first order phase 
transition. The R\'{e}nyi entropy (as defined in Eq. (\ref{Renyi_def}) for $n\rightarrow \infty$) in this case is for $S>2$
\begin{equation}
 \mathcal{S}_{n\rightarrow\infty}=\ln(S+1)+\phi(x)+\phi(1/x),
\end{equation}
with $x=\frac{|\lambda+1|+1}{S+1}$ and
\begin{equation}
 \phi(x)=\beta+\sum_{n=1}^\infty\frac{e^{-n\mu}}{n}{\rm sech}\, n\mu \sin2n\beta.
\end{equation}
Here, $\mu$ and $\beta(x)$ are defined implicitly by the relations $\cosh\mu=(S+1)/2$ and $x=\sinh\mu/\sinh(\mu-\beta)$.

\begin{figure}[th]
		 \includegraphics[width=0.8\linewidth]{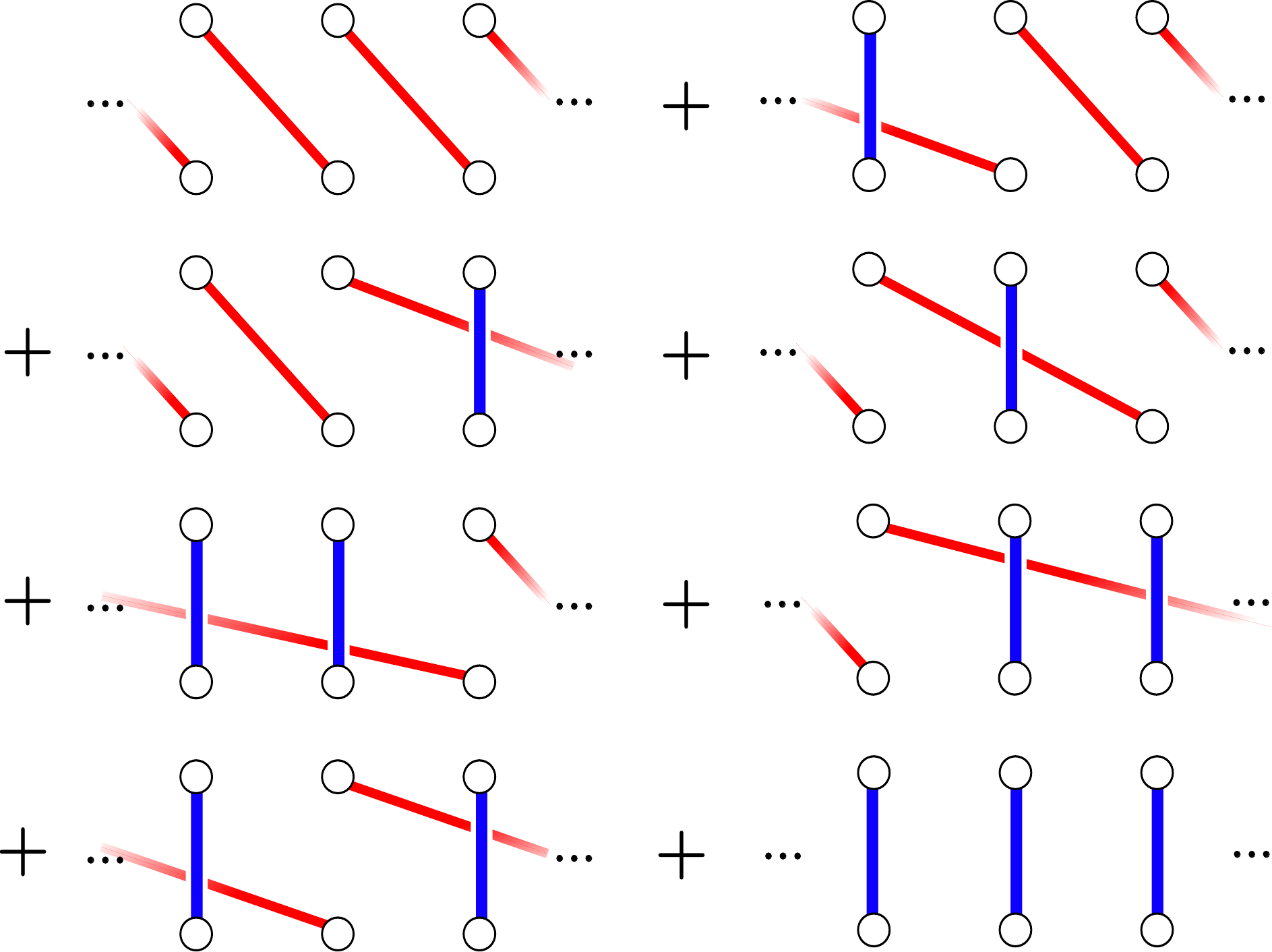}
	\caption{(Color online) All possible dimer coverings of six particles (white circles). Red lines represent 
	singlets between particles at different rungs. Blue lines denote singlets in a rung.
	The MPS, Eq. (\ref{interpolation_exact}), consists of all the possible dimer coverings of singlets between the
	upper and the lower leg of the chain, such that the spin $S/2$ particle at site $i$ in the upper leg
	is connected to the particle of spin $S/2$ at $j\geq i$.}\label{fig:MPS_covering}
\end{figure}

The analog of the latent heat, which we call the \emph{latent entanglement}, in this transition is given by
\begin{equation}
 L=\frac{2}{S+1}\sum_{a=1,2}K_a\exp(K_a)\zeta(x_a)\prod_{m=1}^\infty(\tanh m\mu)^2
\end{equation}
with $x_1=x$ and $x_2=1/x$. Here $\zeta(x)=\sinh\mu/(1+x^2+2x\cosh\mu)$. The appearance of a first order phase transition in the entanglement entropy as a function of the spin might be a signal of the dimerization transition in spin chains with $SU(N)$ symmetry. The appearance of the $SU(N)$ symmetry can be understood looking at the ground state configuration of Eq. (\ref{interpolation_exact}), described in Fig.\ref{fig:MPS_covering}.
This valence bond basis spans the subspace of states that are $SU(N)$ symmetric, with $N=S+1$ (See appendix \ref{app:singlet}).
As discussed in \cite{Affleck1990}, a dimerization transition is expected for antiferromagnets with $SU(N)$. For $S/2=1/2$ the 
ground state of the Heisenberg antiferromagnet is unique as shown from the exact solution. For very large spin $S$ the translation by
one site is spontaneously broken and the system exhibits a gap. A more concrete connection between the latent entanglement
and spontaneously broken symmetries will be studied elsewhere.

\section{Conclusion}\label{sec:CON}

In this article we studied the BES of symmetry protected topologically trivial/nontrivial states in spin ladders. 
We first analyzed a spin 1/2 ladder that generalizes the VBS ground state of the AKLT model. This generalization
is introduced by an MPS description of bond dimension 2. We discuss the mapping of the BEH of this model 
to an XYZ effective spin $1/2$ Hamiltonian. We find for this model that whenever the eigenvalues of the one dimensional transfer 
matrix coincide, the BES becomes critical. This indicates that two BEHs can be adiabatically connected without closing
the entanglement gap as long as their associated one dimensional transfer matrices does not acquire extra degeneracies during 
the interpolation. We showed that BEH of the VBS spin $S$ ground state 
realized in ladders of spin $S/2$ corresponds to an effective spin chain Hamiltonian of spin $S/2$ particles. 
By a direct application of the Haldane conjecture on the BEH, we argue that the BES is gapless for spin $S/2$ ladders realizing the 
Haldane phase with $S$ odd. On the other hand, $S/2$ spin ladders connected to an even spin $S$ VBS ground state possess 
generically an entanglement gap.

Finally, we discussed a non DP deformation of the VBS ground state (for spin ladders) with a finite
correlation length along the deformation, that does change the degeneracies of the one dimensional transfer matrix. This deformation is an illustration of the
non-universality of the BES under non DP deformations. Our work is the first to discuss the non-universality of the BES. The non DP deformation we considered connects the spin $S$ VBS state with a state with an additional 
$SU(N)$ symmetry, with $N=S+1$. The BES of this deformation can be obtained non-perturbatively by mapping the R\'{e}nyi entropy 
of this state to the critical $n$-state Potts model with $n=(S+1)^2$. This may indicate that the Reyni entropy is sensitive to a dimerization transition of the physical ground state, becoming non-analytic for $n\rightarrow \infty$.

\acknowledgments The authors acknowledge the hospitality of the 2014 Les Houches summer school: Topological aspects of condensed matter physics where the beginning stages of this work took place. R.A.S. is supported by GIF. C.M.J. is supported by the David and Lucile Packard foundation. R.L. is supported by National Science Foundation Graduate Fellowship award number 2012115499.

\appendix
%%%%%%%%%%%%%%%%%%%%%
\onecolumngrid
\section{Hamiltonian of generalized spin $S$ VBS on ladders}\label{app:HigherSVBS}
The generalized AKLT model for a spin chain with spin $S$ particles is
\begin{equation}
 \mathcal{H}=\sum_{i=1}^{N_{\rm sites}}\sum_{J=S+1}^{2S}C_JP_J(\hat{S}_j,\hat{S}_{i+1}),
\end{equation}
with $C_J>0$. The operator $P_J(\hat{S}_j,\hat{S}_k)$ is the projector onto the subspace of spin $J$ of the total spin $S_{jk}=\hat{S}_j+\hat{S}_k$.
Explicitly
\begin{equation}
 P_J(\hat{S}_j,\hat{S}_k)=\prod_{\substack{L=0\\L\neq J}}^{S}\frac{(\hat{S}_j+\hat{S}_k)^2-L(L+1)}{J(J+1)-L(L+1)}=
 \prod_{\substack{L=0\\L\neq J}}^{S}\frac{2\hat{S}_{j}\cdot\hat{S}_{k}+S(S/2+1)-L(L+1)}{J(J+1)-L(L+1)}.
\end{equation}
In the case of a spin ladder with particles of spin $S/2$, the above Hamiltonian does not have an unique ground state.
Using the fact that the VBS groundstate of higher spin (as discussed in Sec. \ref{sec:spinSladder}) 
can be constructed locally by forming singlets between nearest neighbors and symmetrizing the spins two spin $S/2$ in the
rungs onto a spin $S$, the VBS state on the ladder minimizes the energy of the following Hamiltonian,
\begin{equation}
 H=-\sum_{i=1}^{N_{\rm sites}} P_S(\hat{S}_{i,1},\hat{S}_{i,2})P_{0}(\hat{S}_{i,1},\hat{S}_{i+1,2})P_S(\hat{S}_{i,1},\hat{S}_{i,2}).
\end{equation}
%%%%%%%%%%%%%%%%%
\section{Transfer matrix for general spin $S$ state}\label{appendixB}
\begin{figure}[h!]
		 \includegraphics[width=0.5\linewidth]{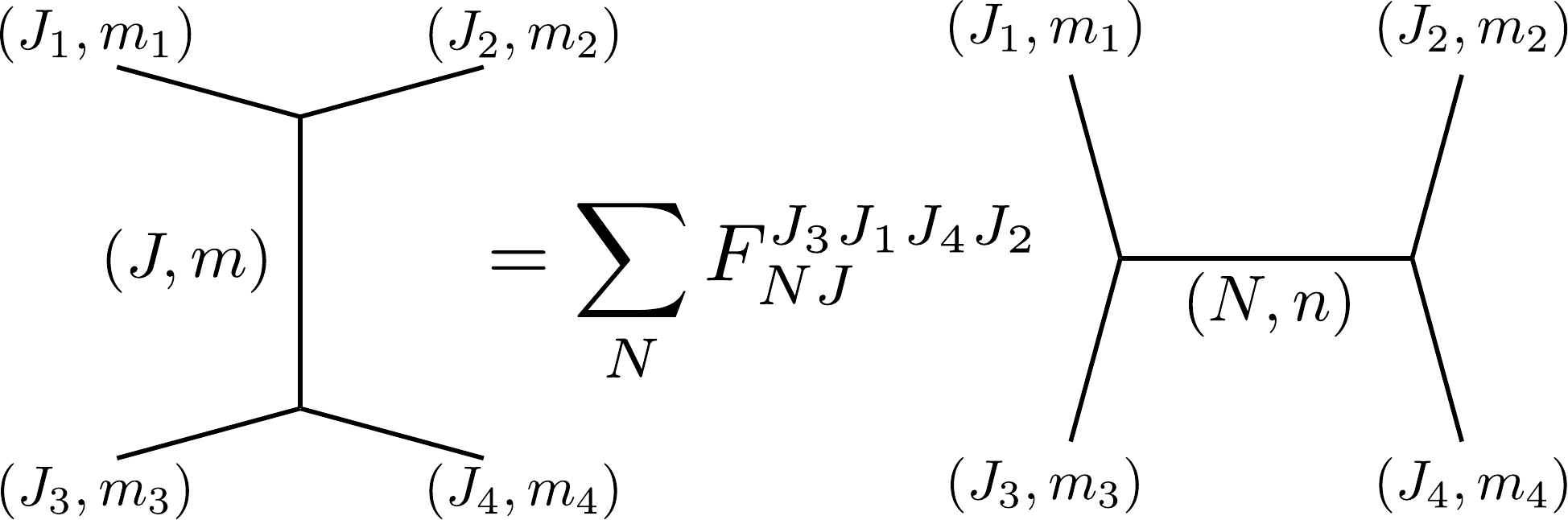}
	\caption{Recoupling of Clebsch-Gordan coefficients.}\label{fig:F_matrix}
\end{figure}

%%%%%%%%%%%%%%%%%
In order to compute the eigenvalues of the transfer matrix $\tilde{T}$, we make use of the SU($2$) structure of $\tilde{T}$. This 
matrix corresponds to the contraction of Clebsch-Gordan symbols, so we can use the $F-$matrix (Racah coefficients 
\cite{messiah1965quantum}) to recouple the coefficients. The recoupling is expressed in Fig. (\ref{fig:F_matrix}) and algebraically it reads
\begin{eqnarray}\nonumber
 \sum_m\ClebschG{J_1}{m_1}{J_2}{m_2}{J}{m}\ClebschG{J_3}{m_3}{J_4}{m_4}{J}{m}\frac{(-1)^{J_2-J+m_1}}{\sqrt{2J+1}}=
 \sum_{N,n}F_{NJ}^{J_3J_1J_4J_2}\ClebschG{J_1}{-m_1}{J_3}{m_3}{N}{n}\ClebschG{J_2}{m_2}{J_4}{-m_4}{N}{n}\frac{(-1)^{N-J_2+m_4}}{\sqrt{2N+1}}.
\end{eqnarray}

The $F-$matrix is related to the $6j$-symbol by \cite{messiah1965quantum}
\begin{equation}
 F^{J_1J_2J_3J_4}_{NJ}=(-1)^{J_1+J_2+J_3+J_4}\sqrt{(2N+1)(2J+1)}\SixJun{J_1}{J_4}{J_2}{J_3}{N}{J}.
\end{equation}

%%%%%%%%%%%%%%%%%
\section{Large $S$ limit of $6j$ symbol}\label{app:6j}
%%%%%%%%%%%%%%%%%

In Sec. \ref{sec:spinSladder}, we obtained the BEH (Eq. (\ref{BEH_antiferro})), whose coefficients are given in terms of the $6j$ symbols.
Explicitly, in terms of the Beta function ${\rm B}(x,y)$,
\begin{equation}
 \SixJun{S}{\frac{S}{2}}{S}{\frac{S}{2}}{j}{\frac{S}{2}}^2(S+1)^3=\frac{(S+1)^2}{2S+1}\frac{{{\rm B}}(S+2,S+1)}{{\rm B}(2S+1,2S+2)}
 \frac{{\rm B}(2S-j+1,2S+j+2)}{{\rm B}(S+j+2,S-j+1)}.
\end{equation}
Using the relation between the Beta function and the Binomial coefficient ${\rm B}(a,b)=(b+1)^{-1}\binom{a+b-1}{b}^{-1}$, together
with the approximation of the Binomial coefficient
\begin{equation}
 \binom{n}{k} = \frac{2^{n+1/2}}{\sqrt{n \pi }} e^{-\frac{(k-(n/2))^2}{n/2}}\left[1+\mathcal{O}\left(\frac{1}{\sqrt{n}}\right)\right],
\end{equation}
we have
\begin{equation}
 \SixJun{S}{\frac{S}{2}}{S}{\frac{S}{2}}{j}{\frac{S}{2}}^2(S+1)^3=\frac{1}{4^S}\sqrt{\frac{(S+1)^3}{2S+1}}\frac{{{\rm B}}(S+2,S+1)}{{\rm B}(2S+1,2S+2)}
 \frac{S-j+1}{2S+j+2}\exp\left(-\frac{Sj^2}{(S+1)(2S+1)}\right)\left[1+\mathcal{O}\left(\frac{1}{\sqrt{S}}\right)\right].
\end{equation}

%%%%%%%%%%%%%%
\section{SU($N$) symmetry of singlet states}\label{app:singlet}
%%%%%%%%%%%%%%%%%
The singlet state, written explicitly in terms of the $S_z$ basis states, is
\begin{equation}
 |\phi\rangle=\frac{1}{\sqrt{S+1}}\sum_{m=-S/2}^{S/2}(-1)^{S-m}|S/2,m\rangle\otimes|S/2,-m\rangle,
\end{equation}
which can be compactly written as
\begin{equation}
 |\phi\rangle=\frac{1}{\sqrt{S+1}}\sum_{m=-S/2}^{S/2}|S/2,m\rangle\otimes\mathcal{T}|S/2,m\rangle,
\end{equation}
by using the time reversal operator, $\mathcal{T}$. Changing the basis by a unitary transformation (by means
of a $(S+1)\times(S+1)$ matrix $U$, such that $|a\rangle=\sum_mU_{am}|S/2,m\rangle$) we obtain
\begin{equation}
 \frac{1}{\sqrt{S+1}}\sum_{a}|a\rangle\otimes\mathcal{T}|a\rangle=\frac{1}{\sqrt{S+1}}\sum_{a,m,m'}U_{am}U_{am}^*|S/2,m\rangle\otimes\mathcal{T}|S/2,m'\rangle
=\frac{1}{\sqrt{S+1}}\sum_{m=-S/2}^{S/2}|S/2,m\rangle\otimes\mathcal{T}|S/2,m\rangle,
 \end{equation}
where the unitarity of $U$ was used. This implies that the $SU(2)$ singlet is also a SU($N$) singlet, with $N=S+1$.
\twocolumngrid

%\bibliography{bib.bib}
%Merlin.mbs v4.21 2009-07-09.
%

\end{document}